\documentclass[11pt,a4paper]{article}
\usepackage{acronym}
\usepackage{algorithm}
\usepackage{algorithmic}
\usepackage{subcaption} 
\usepackage{slashbox}
\usepackage{amssymb}
\usepackage{amsmath}
\usepackage{multicol}
\usepackage{graphicx}
\usepackage[hyphens]{url}
\acrodef{CNN}{Convolutional Neural Network}
\acrodef{FPGA}{Field-Programmable Gate Array}
\acrodef{CPU}{Central Processing Unit}
\acrodef{GPU}{Graphics Processing Unit}
\acrodef{GPGPU}{General-Purpose GPU}
\acrodef{DNN}{Deep Neural Network}
\acrodef{RAM}{Random-Access Memory}
\acrodef{OpenCL}{Open Computing Language}
\acrodef{CUDA}{Compute Unified Device Architecture}
\acrodef{IP}{Internet Protocol}
\acrodef{ReLU}{Rectified Linear Unit}
\acrodef{FLOPs}{Floating-Point Operations per second}
\acrodef{SIMT}{Single Instruction, Multiple Thread}

\usepackage[affil-it]{authblk}

\begin{document}



\title{Distributed learning of CNNs on heterogeneous CPU/GPU architectures}

\author{Jose Marques}
\affil{Instituto de Telecomunica\c{c}\~oes, Dept. of Electrical and Computer Engineering, University of Coimbra, P\'{o}lo II - Universidade de Coimbra, 3030-290 Coimbra, Portugal}

\author{Gabriel Falcao}
\affil{Instituto de Telecomunica\c{c}\~oes, Dept. of Electrical and Computer Engineering, University of Coimbra, P\'{o}lo II - Universidade de Coimbra, 3030-290 Coimbra, Portugal}

\author{Lu\'{i}s~A.~Alexandre}
\affil{Instituto de Telecomunica\c{c}\~{o}es, Universidade da Beira Interior, Rua Marqu\^es d'\'Avila e Bolama, 6201-001, Covilh\~a, Portugal
}

\maketitle

%
%
%
%
%
%

\begin{abstract}

Convolutional Neural Networks (CNNs) have shown to be powerful classification tools in tasks that range from check reading to medical diagnosis, reaching close to human perception, and in some cases surpassing it. However, the problems to solve are becoming larger and more complex, which translates to larger CNNs, leading to longer training times---the computational complex part---that not even the adoption of Graphics Processing Units (GPUs) could keep up to. This problem is partially solved by using more processing units and distributed training methods that are offered by several frameworks dedicated to neural network training, such as Caffe, Torch or TensorFlow. However, these techniques do not take full advantage of the possible parallelization offered by CNNs and the cooperative use of heterogeneous devices with different processing capabilities, clock speeds, memory size, among others. This paper presents a new method for the parallel training of CNNs that can be considered as a particular instantiation of model parallelism, where only the convolutional layer is distributed. In fact, the convolutions processed during training (forward and backward propagation included) represent from $60$-$90$\% of global processing time. The paper analyzes the influence of network size, bandwidth, batch size, number of devices, including their processing capabilities, and other parameters. Results show that this technique is capable of diminishing the training time without affecting the classification performance for both CPUs and GPUs. For the CIFAR-10 dataset, using a CNN with two convolutional layers, and $500$ and $1500$ kernels, respectively, best speedups achieve $3.28\times$ using four CPUs and $2.45\times$ with three GPUs. Modern imaging datasets, larger and more complex than CIFAR-10 will certainly require more than $60$-$90$\% of processing time calculating convolutions, and speedups will tend to increase accordingly.

\end{abstract}


\section{Introduction}
\label{intro}

Deep learning has been the engine behind tasks that are considered common nowadays, such as using search engines, or depositing checks at the ATM~\cite{lecun1997reading}, including filtering content on social media, or even tasks that range from medical diagnosis~\cite{kononenko2001machine} to game playing~\cite{mnih2013atari}. It is ever more present all around, particularly on smart appliances (like smart homes~\cite{kabir2015machine,dixit2014use} and smartphones~\cite{falcao2016evaluation}).

One of the most used models within deep learning is the \ac{CNN}. The first major contribution from a \ac{CNN} to the growth of deep learning appeared when this kind of network was used to win the ILSVRC, the largest contest in object recognition, by diminishing the top-5 error rate from 26.1\% to 15.3\%~\cite{krizhevsky2012imagenet}. The \ac{CNN} creates a list of possible categories for each image, from 1000 possible categories, and the correct one always appears amongst the first 5 except 15.3\% of the time.

However, with the boom of deep learning in general, and \acp{CNN} in specific, came the increase of the samples per dataset as well as an increase in models' size. Where in the 1990s the datasets had less than a couple thousand instances available to train, a decade later the creation of datasets spiked, as well as the number of instances they contain. The same tendency happened to the networks' size, with the first deep networks having few neurons while most recent networks can have from millions to billions of parameters.

The technological development allowed the access to computational resources capable of training increasingly larger neural networks, and also larger datasets. More specifically, it was due to the development of faster \acp{CPU} and \acp{RAM}, the increase in available memory/ storage, and also due to the improvement of distributed training infrastructures. It was only then that it was possible to create frameworks like DistBelief~\cite{corrado2012large}, capable of training networks with as much as 1.7 billion parameters, currently among the largest of their type. However, it should be noted that this framework uses thousands of \ac{CPU} cores distributed along hundreds of machines and the training takes days to complete. 

Another important technological development for the evolution of deep learning was the adoption of \ac{GPU} architectures, more specifically the use of \ac{GPGPU} for scientific and generic computation. These newly available resources allowed speeding up network training, through the development of parallel computing frameworks like \ac{OpenCL}~\cite{OpenCL} and \ac{CUDA}~\cite{CUDA}. This is only possible because \acp{GPU}, despite working at smaller frequencies than \acp{CPU}, have a higher number of cores that are very efficient at receiving a large batch of data and repeating the same operation very quickly using a \ac{SIMT} programming model, something that happens recurrently during neural network training. 

A distinctive aspect of \acp{CNN} lies on the so called convolutional layers, which makes it the ideal choice for image and speech recognition, since both tasks rely heavily on the correlation of neighbouring data. The major problem with it is that the processing of convolutions is computationally intensive, requiring from 60\% to 90\% of the total training time only using about 5\% of the parameters of the whole network~\cite{ward2011efficient,krizhevsky2014weird}. Although for such scenario, Amdahl's Law~\cite{gene2013computer} constrains speedups to the range $2.5 \sim 10$, this work shows that it is possible to work near those limits. Also, while tools such as Caffe or TensorFlow usually explore the compute power of a single GPU or a small group of homogeneous GPUs on the same node / server, this work proposes an alternative for filling in the gaps of the above mentioned frameworks, namely by using truly heterogeneous CPU and GPU parallel computing architectures, isolated or grouped in distinct nodes / clusters, eventually in different physical locations, for the compute-intensive training of deep learning with balanced workloads.

Thus, the main contribution of this paper consists in an open source distribution technique that makes use of the potential parallelization that convolutional layers have to offer, feeding platforms of conventional heterogeneous CPU and GPU devices the same feature maps, but providing them with different kernels and balanced workloads, gaining speed up during the computation of convolutions that compensates communication times for orchestrating the different nodes. This contribution is likely to produce significant impact, since under the current context of training CNNs, computation times can easily vary from days to weeks~\cite{keuper2016distributed}, or even more, until classification errors converge to the desired small level targets.

\section{Distributed Convolutional Neural Networks}

\label{chapter:cnn}
Over the last years, \acp{DNN} have shown great promise in several practical applications, achieving state-of-the-art performance on a variety of different tasks, including object recognition~\cite{erhan2013scalable,szegedy2013deep} and speech recognition~\cite{toth2014combining,abdelhamid2014convolutional}, among others. They were able to achieve strong super-human performances, performing better than all humans or the best ones at it, in games like chess~\cite{lai2015giraffe} and Go~\cite{silver2016mastering}.

Despite having shown to be a powerful machine learning technique, when applied to large inputs, like images, most \acp{DNN} like deep belief nets or stacked autoencoders, became rather complex and sizable, capable of reaching millions of weights for simple inputs that consist of RGB images of size $32\times32$.

Another problem is that these networks neglect correlation between neighbouring data, like translations and distortions, despite there are local correlations in pattern recognition problems. Ideally, local features would be extracted and analyzed in order to be able to detect certain beings or objects. \acp{CNN}, however, are able to overcome those issues by making use of $3$ key factors: local receptive fields, weight sharing and spatial pooling.

\subsection{Convolutional Neural Networks}

The first proposal of a model similar to a \ac{CNN} can be attributed to Fukushima with the neocognitron, in 1980~\cite{Fukushima1980}. It served as inspiration to the modern concept of \ac{CNN}, introduced in 1995 by Yann LeCun and Yoshua Bengio \cite{lecun1995convolutional}, also inspired by the discovery of locally sensitive and orientation-selective neurons in the visual cortex of a cat. By using local receptive fields it is possible to exploit local visual features, like edges, corners and end-points (in images). This is advantageous because adjacent pixels tend to be strongly correlated while pixels that are farther apart are usually uncorrelated, or have weak correlation. Having the ability to share weights across locally connected neurons allows reducing the amount of parameters to train, decreasing the amount of data needed, making the training faster and achieving better classification performances when compared to other approaches.

\begin{center}
\begin{figure}[h!]
\includegraphics[width=\columnwidth]{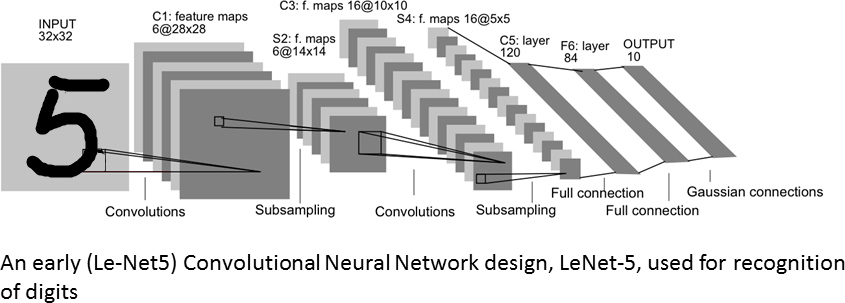}
\caption{The architecture of LeNet-5~\cite{lecun1989backpropagation}, a \ac{CNN} used for digits recognition for the MNIST dataset.
\label{fig:convolutionalnetwork}}
\end{figure}
\end{center}

The main differences between \acp{CNN} and other \acp{DNN} are the use of convolutions and pooling (or subsampling) operations, instead of simple matrix multiplication in at least one layer. One of the most popular \acp{CNN} is LeNet-5~\cite{lecun1989backpropagation}, illustrated in Figure \ref{fig:convolutionalnetwork}, for digits recognition. It contains 2 convolutional layers and 2 subsampling layers interleaved, ending with fully connected layers. The network can be tuned by changing the number of layers, the number and size of filters from the convolutional layer or the stride of the subsampling layer.

\subsubsection{Convolutional Layer}

The input of a convolutional layer is usually a multidimensional array of data, while the kernel (or filter) is a multidimensional array of parameters that readjusts through the network training. A convolution operation then applies those kernels to the inputs, as to detect the most appropriate features. 

The reason for the popularity of convolutional layers is due to their ability to work with variable sized inputs, to which sparse connectivity and parameter sharing provided important contributions.

Sparse connectivity (or sparse weights) occurs when the outputs only have a reduced amount of connections. Considering the case of an image, that may have thousands to millions of pixels, a kernel is used to detect small features, thus storing few parameters and limiting the number of outputs.

Parameter sharing is used to reduce the number of parameters, which can be achieved using the same filter across the entire input. This means that instead of learning a separate set of parameters for every possible location, it is only necessary to learn one set, allowing the detection of features regardless of their position in the input. To learn more features, more filters must be used, so that they can be trained to detect different features.

\subsubsection{Pooling Layer}

The pooling layer is a form of down-sampling, that partitions its input into several non-overlapping blocks and evaluates a pooling function over each block. 

In every pooling function, the goal is to make the network invariant to small transformations, meaning that if the input was translated by a small amount, the values of most pooled outputs would remain the same, which is particularly important if the presence of a certain feature is more relevant than its position. 

The pooling layer can also be used to perform dimensionality reduction in the feature map, trimming the amount of parameters and computation required to train the network, thus controlling the overfitting.

Convolutional and pooling layers represent the major differences between \acp{DNN} and \acp{CNN}, but this type of neural networks makes use of several other layers, particularly \ac{ReLU}, fully connected and loss layers.

\subsection{Distributed Training Techniques}

Training the largest \acp{CNN} is becoming a real challenge even using \acp{GPU}, either because datasets are growing fast in size and these parallel machines are limited in memory, or simply because the training times still remain quite long. Performing the distributed training of \acp{CNN} fosters accelerating this type of complex processing. This section aims to provide some insight regarding the most recent techniques of distributed training.

Distributed training can refer to distributing the training of the network across several \acp{GPU} or CPUS in the same computer or in different computers. There are mainly two types of techniques for performing the distribution: data parallelism and model parallelism.

\subsubsection{Data Parallelism}

In data parallelism the batch of data is split across the several nodes of the cluster, may they be \acp{CPU}, \acp{GPU} or a combination of both. Each node is then responsible for computing the gradients with respect to all the parameters, but does so using part of the batch. However, since every node is running a replica, it is necessary to communicate the gradients and parameter values on every update step. Another problem with this approach is that since every node calculates different gradients, they need to be averaged, and that causes the loss of information and may hinder the training process.

\begin{center}
\begin{figure}[h!]
		\includegraphics[width=\columnwidth]{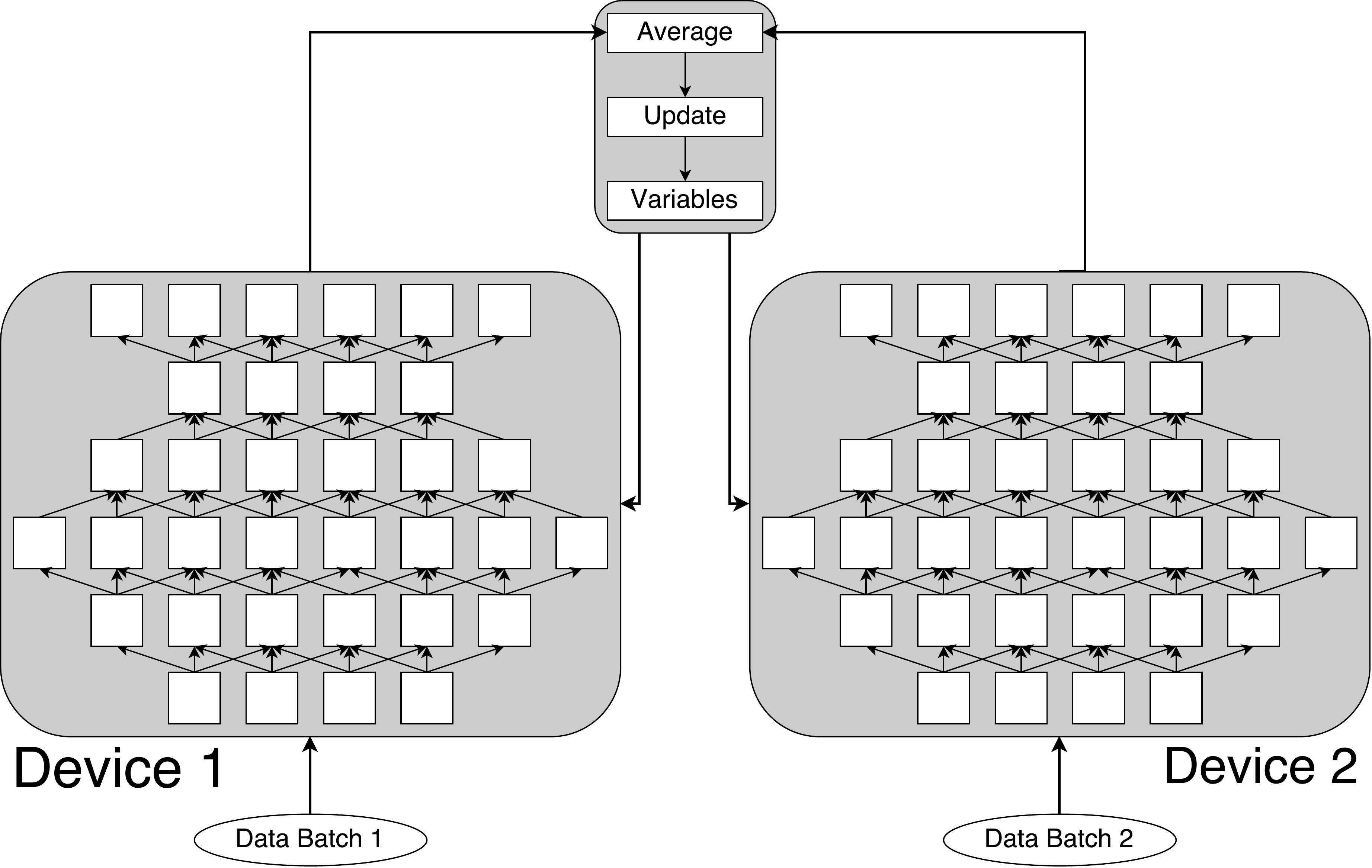}
		\caption{Data parallelism. Each slave node receives a different batch of data and trains a model replica, also computing the gradients. The gradients of every slave node are sent to the master node and averaged, with the parameters being updated across all replicas.}
		\label{fig:dataparallelism}
\end{figure}
\end{center}

Another condition for the use of this type of parallelism, especially when using \acp{GPU} is that the batch size must be large enough to be distributed and still be able to exploit the highly parallel capabilities of the \ac{GPU}. An example of data parallelism is presented in Figure \ref{fig:dataparallelism}.

\subsubsection{Model Parallelism}

\begin{center}
\begin{figure}[h!]
		\includegraphics[width=\columnwidth]{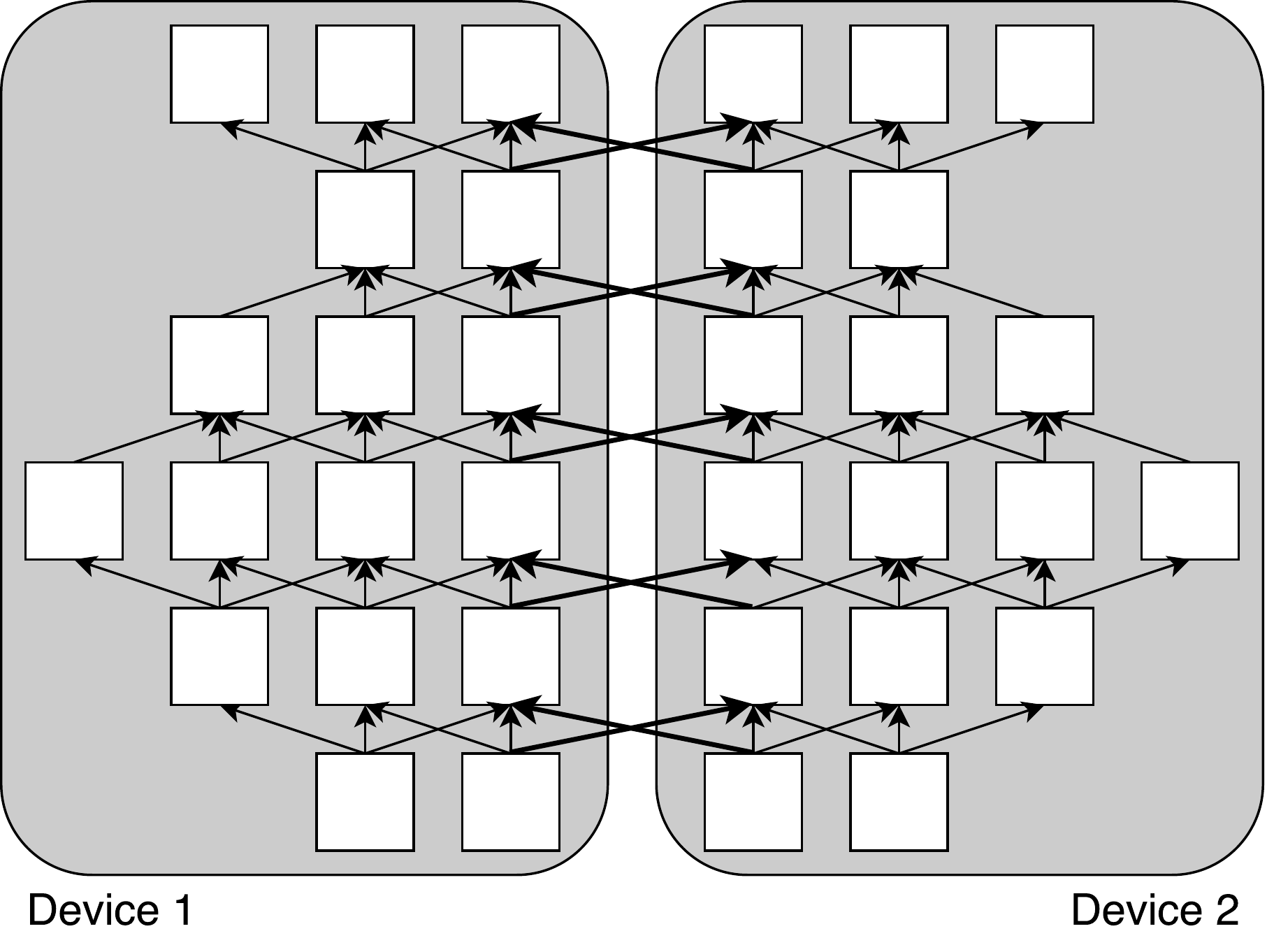}
		\caption{Model parallelism. The network is divided across all devices, with nodes having edges that cross partition boundaries transmitting their state between devices.}
		\label{fig:modelparallelism}
\end{figure}
\end{center}

Model parallelism consists of dividing the network's computation across the several nodes, that may differ considering the type of network used. In the DistBelief~\cite{corrado2012large} case, the \ac{DNN} is partitioned across several nodes and only the nodes with edges that cross partition boundaries need to have their state transmitted between nodes. Another possible implementation~\cite{ranzato2013multi} separates the first convolutional layer across several nodes, dividing the number of kernels, with each node calculating a part of the network, having only cross connections at one intermediate layer and at the very top fully connected layers. This implementation is visible in Figure \ref{fig:modelparallelism}.

A different type of model parallelism can also be considered by splitting the image in tiles that are represented by thread blocks per output feature map. Each tile is analogous to a thread block and each pixel is represented by a thread, with a tile representing a different image~\cite{ward2011efficient}. However, this type of distribution is only efficient in cases where the image and batch size is large enough, and when there are not many kernels to be convoluted, since every device will need to have every kernel.

The distribution technique devised in this paper can be thought as a new type of model parallelism, since the workload of the network is distributed across several machines. This will be further detailed in Section \ref{chapter:methodology}.

\section{Related Work}

There are a few works that address the speedup achievable with distribution techniques, mostly data parallel ones. Among the few frameworks that allow the distributed training of convolutional networks, TensorFlow~\cite{tensorflow2015} offers the possibility of training a model in a parallel, distributed fashion, providing the code to do so with the CIFAR-10 dataset~\cite{krizhevsky2009learning}, using several GPUs on the same machine. Each GPU is envisioned to have similar speed and have enough memory for the entire network, so a model replica is placed in each GPU, with the model parameters being updated synchronously, having to wait for all GPUs to complete the processing of the corresponding data.

The results provided by TensorFlow~\cite{multigpu}, commented in file \textit{cifar10\_multi\_gpu\_train.py} of the source code, are as follows in Table \ref{table:tensorflow}.
\begin{table}
	\centering
	\begin{tabular}{lll}
		\hline\noalign{\smallskip}
		System & Step Time  & Accuracy \\
		 & (s/batch) & \\
		\noalign{\smallskip}\hline\noalign{\smallskip}
		1 Tesla K20M & 0.35-0.60 & \textasciitilde86\%, 60K steps \\
		2 Tesla K20M & 0.13-0.20 & \textasciitilde84\%, 30K steps \\
		3 Tesla K20M & 0.13-0.18 & \textasciitilde84\%, 30K steps \\
		4 Tesla K20M & \textasciitilde0.10 & \textasciitilde84\%, 30K steps \\
		\noalign{\smallskip}\hline
	\end{tabular}
	\caption{Results using TensorFlow multi-GPU training for the CIFAR-10 dataset, using the test images~\cite{multigpu}.}
	\label{table:tensorflow}
\end{table}
As the table shows, the introduction of a second GPU is able to reduce the step time to less than half of the case with a single device. However, for the remaining cases, with 3 and 4 GPUs, the step time is barely reduced, showing that it doesn't seem to be scalable.

There are also studies that try to distribute the training of a CNN across different machines~\cite{vishnu2016distributed}. Vishnu \textit{et al.} use several CPUs connected using InfiniBand. For the implementation, the framework used was also TensorFlow, taking advantage of data parallelism. All the samples are divided equally across devices, as each device is considered to have the same exact computational capabilities. The updates are performed synchronously using MPI, which it is heavily optimized, allowing for a minimal time being spent in communications. The results relative to a CNN trained with CIFAR-10 show a speedup of $3.01\times$ when scaling from $4$ to $64$ cores.

Also, the fact that distinct devices may receive the same amount of data and update the parameters synchronously are limitations to the use of real life machines that have different capabilities. This is the main motivation in developing a significantly distinct approach capable of performing CNN training in truly heterogeneous devices. Another problem is the fact that these studies seem to be limited to distribute the training across several devices (e.g. GPUs) placed in a single machine.
	
Thus, our proposal is to develop a parallelization scheme that is able to train a CNN using the resources of different machines available on a network, with distinct computational resources. To avoid limiting the training time to the slowest machine used, a quick test is performed on all machines, so as to grasp the computational capabilities of each device. Since the distribution of the workload is performed during runtime, it allows the use of a wide variety of devices, each one receiving a proportional share of the workload.

\section{Distributed Convolutional Learning}
\label{chapter:methodology}

The goal is to obtain a method of distributed training that takes advantage of the structure of a \ac{CNN}. As stated before, the convolutional phase makes up for at least 60\% of the training time~\cite{ward2011efficient} and may even represent up to 90\% of computation time, using from 5\% to 10\% of the network's parameters~\cite{krizhevsky2014weird}.

\subsection{Problem analysis}

The first step towards implementation lies in the analysis and definition of the task at hand. The current approach is a variant of the model parallelism where only the convolutional layer is distributed. A more detailed description follows in Sections \ref{subsec:hybrid} and \ref{section:distributed}.

\subsubsection{Hybrid CPU-CPU and GPU-GPU computing}
\label{subsec:hybrid}

One of the major problems that arises with the usage of computers having different \acp{CPU} and \acp{GPU} is that different devices have different computational resources and thus are able to complete the same workload in different times. This can become a problem, especially when one or several devices are relatively slow compared to the others. For example, considering Device 1, that can complete an arbitrary workload in only 10 seconds, and Device 2, that completes the same work in 20 seconds, if the workload were to be distributed equally, Device 1 would complete the task in 5 seconds while Device 2 would take 10 seconds. If Device 1 were to be used as the comparison basis, the speed up would be below 1x, as computation time would remain the same, but communication times would be introduced.

In order to mitigate this problem, it is necessary to find beforehand the suitable workload for each device, which in this case is the number of kernels, so that each device can finish all its convolutions at approximately the same time.

To do so, a pre-processing procedure is performed, where every device runs a N-dimensional convolution with both the size of the images and the size of the kernels provided by the master device, trying to simulate part of the convolutional layer. The convolution is run using random values, since only the time spent performing calculations is relevant. After the respective simulations complete, the computation time is reported to the master node in order to find the performance ratio between devices, either \acp{CPU} or \acp{GPU}. The slave nodes only need to know the \ac{IP} address of the master node, while the master node needs to know the number of slave nodes and their respective IP addresses.

Considering the same example as before, the performance values would be [2, 1], for Device 1 and Device 2, respectively. Device 1 would then have twice as much kernels as Device 2, for the same arbitrary workload, with Device 1 in charge of two thirds and Device 2 convolving only one third of the kernels. This means that both devices would finish their convolutions in about 6.67 seconds, which taking into consideration the previous processing time of 10 seconds represents a speed up of 1.5x. This difference in performance between devices comes from the differences regarding data transfers and computing capabilities.

However, considering that during the experiment three to four computers will be used, it is necessary to further clarify how the attribution of performance values and subsequent distribution of work is done, in cases with more than two devices. In general, for $n$ devices, each with a time to complete the task given by $t_i, \ i = 1, \ldots,n$, we define the workload for each device as:
\begin{center}
	\begin{equation}
	w_i = \frac{\frac{\max(t)}{t_i}}{\sum_{j=1}^{n}\frac{\max(t)}{t_j}} .
	\end{equation}
\end{center}

\subsubsection{Workload distribution}
\label{section:distributed}

\begin{algorithm}[h!]
	\caption{Master Node}
	\label{alg:master}
	{\footnotesize
	\begin{algorithmic}[1]
		\FOR {$slave = 1$ to $numSlaves$}
		\STATE {$connectSocket(slave)$}
		\ENDFOR
		\STATE
		\WHILE {training}
		\FOR {$layer = 1$ to $numLayers$}
		\IF {convolutional layer}
		\FOR {$slave = 1$ to $numSlaves$}
		\STATE
		\COMMENT {All slaves receive same inputs but different kernels.}
		\STATE {$inputs \Rightarrow writeSocket(slave)$}
		\STATE {$numMaps(slave) \Rightarrow writeSocket(slave)$}
		\STATE {$kernels(slave) \Rightarrow writeSocket(slave)$}
		\ENDFOR
		\STATE
		\FOR {$maps = 1$ to $numMaps$}
		\STATE {$output = convn(inputs, maps)$}
		\ENDFOR
		\STATE
		\COMMENT {Master node receives all feature maps.}
		\FOR {$slave = 1$ to $numSlaves$}
		\STATE {$output(slave) \Leftarrow readSocket(slave)$}
		\STATE {$allOk \Rightarrow writeSocket(slave)$}
		\ENDFOR
		\ENDIF
		\ENDFOR
		\ENDWHILE
		\STATE
		\FOR {$slave = 1$ to $numSlaves$}
		\STATE {$trainOver \Rightarrow writeSocket(slave)$}
		\ENDFOR
	\end{algorithmic}
	}
\end{algorithm}

The similarities between this distributed approach and model parallelism lie in sharing part of the network. However, where the nodes using model parallelism always compute the same part of the network and communications are kept to a minimum, this approach only calculates convolutions for the convolutional layer. This exploits the fact that convolutional layers use less than 10\% of the parameters~\cite{krizhevsky2014weird}, indicating that communication overheads will not become a relevant problem when compared to the computation time saved.

\begin{algorithm}[h!]
	\caption{Slave Node}
	\label{alg:slave}
	{\footnotesize
	\begin{algorithmic}[1]
		\STATE {$connectSocket(server)$}
		\STATE
		\COMMENT {When the training is over, the master activates a flag that tells the slaves to shutdown.}
		\WHILE {$trainOver = 0$}
		\WHILE {$bytesReceived = 0$}
		\STATE {pause(1)}
		\ENDWHILE
		\STATE
		\COMMENT {The slave nodes receive the inputs, kernels and number of feature maps that it should produce.}
		\STATE {$inputs \Leftarrow readSocket(server)$}
		\STATE {$numMaps \Leftarrow readSocket(server)$}
		\STATE {$kernels \Leftarrow readSocket(server)$}
		\STATE
		\FOR {$maps = 1$ to $numMaps$}
		\STATE {$output = convn(inputs, maps)$}
		\ENDFOR
		\STATE
		\STATE {$output \Rightarrow writeSocket(server)$}
		\STATE
		\COMMENT {After every batch, the server sends an acknowledgement flag.}
		\STATE {$allOk \Leftarrow readSocket(server)$}
		\ENDWHILE
		\STATE
		\FOR {$slave = 1$ to $numSlaves$}
		\STATE {$trainOver \Rightarrow writeSocket(slave)$}
		\ENDFOR
	\end{algorithmic}
	}
\end{algorithm}

For this approach, one of the nodes orchestrates and is designated as master node, while the remaining ones are slave nodes. Considering only the convolutional layer is the subject of distributed training, the master node is in charge of training the remaining network.

\begin{figure*}[!b]
	\begin{center}
		\includegraphics[width=0.8\textwidth]{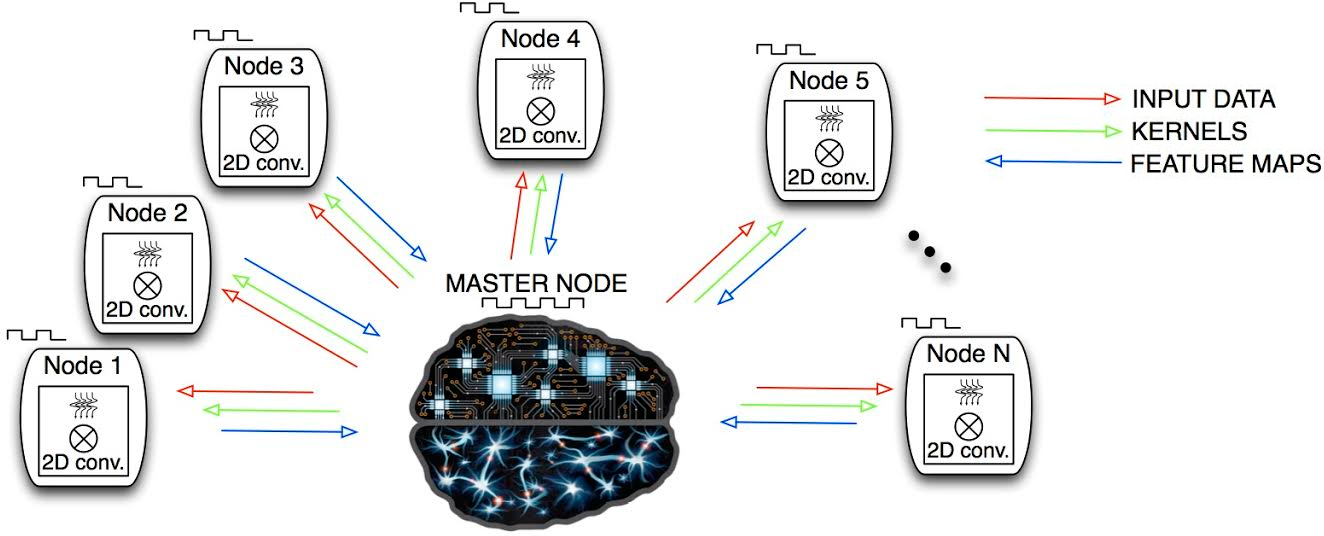}
		\caption{Visualization of computation and information distribution between processing nodes.}
		\label{fig:nodes}
	\end{center}
\end{figure*}
For the convolution distribution, the master node sends the size and number of inputs, that can be images or feature maps from previous layers. It also sends the size and number of kernels needed for the convolution, with different nodes receiving a different number of kernels. All this information regarding input, kernel size and number is necessary so that the slave knows how much data to read from the socket and how it should reshape it, since data read from sockets comes in vector form. After every node concludes their part of the convolutions, each slave sends the feature maps, after which the master node reshapes and rearranges them. The distribution of information between nodes can be further illustrated in Figure \ref{fig:nodes}.

The process is repeated until the training of the network is over, with the master node sending a shutdown flag to every slave. This training distribution technique can be better evaluated by analyzing Algorithms \ref{alg:master} and \ref{alg:slave}, referring to the master and slave nodes, respectively.

\section{Experiments}
\subsection{Hardware platforms setup}
\label{subsec:hardware}

This section discloses the computer platforms' specifications used to conduct the experiment, in Tables \ref{table:cpu} and \ref{table:gpu}, for \ac{CPU} and \ac{GPU} respectively.

\begin{table}[!h]
	\centering
	\begin{tabular}{lll}
		\hline\noalign{\smallskip}
		& Device & RAM \\
		\noalign{\smallskip}\hline\noalign{\smallskip}
		PC1 & Intel i5-3210M @ 2.5 GHz  & 6GB \\
		PC2 & Intel i7-4700HQ @ 2.4 GHz  & 8GB  \\
		PC3 & Intel i7-5500U @ 2.4 GHz  & 8GB \\
		PC4 & Intel i7-6700HQ @ 2.6 GHz  & 16GB \\
		\noalign{\smallskip}\hline
	\end{tabular}
	\caption{CPUs used during the experiments.}
	\label{table:cpu}
\end{table}

As it can be noticed from Tables \ref{table:cpu} and \ref{table:gpu}, the computers are all composed by a set of distinct devices, so the need to perform hybrid \ac{CPU}-\ac{CPU} and \ac{GPU}-\ac{GPU} processing has emerged. The code from this experiment was all written in Matlab. Thus the compatible framework for performing parallel computing is \ac{CUDA}, meaning that the \ac{GPU} cluster uses 3 computers (PC2, PC3 and PC4), since only NVIDIA \acp{GPU} are supported. The \ac{CPU} cluster runs with all computers available.

\begin{table}[!h]
	\centering
	\begin{tabular}{lll}
		\hline\noalign{\smallskip}
		& Device & RAM \\
		\noalign{\smallskip}\hline\noalign{\smallskip}
		PC1 & Radeon HD 7500M & N/A \\
		PC2 & NVIDIA GeForce 840M & 2GB  \\
		PC3 & NVIDIA GeForce 940M & 2GB \\
		PC4 & NVIDIA GeForce GTX 950M & 4GB \\
		\noalign{\smallskip}\hline
	\end{tabular}
	\caption{GPUs used during the experiments.}
	\label{table:gpu}
\end{table}

\subsection{Network architecture and dataset}

For this experiment, the dataset used was CIFAR-10~\cite{krizhevsky2009learning}. It consists of a labeled subset of the 80 million tiny images dataset in~\cite{torralba2008tiny} and was collected by Alex Krizhevsky, Vinod Nair and Geoffrey Hinton. The dataset contains $60000$ $32\times32$ colour images grouped into 10 classes, with each class having $6000$ images. Of the total $60000$ images, $50000$ are intended to be used for training and the remaining $10000$ for testing. The classes present in this dataset are: airplane, automobile, bird, cat, deer, dog, frog, horse, ship and truck. This dataset was chosen particularly for consisting of colour images, which is the norm for most recent image datasets, but also for having a considerably small dataset with small images, which allows to test several \ac{CNN} architectures in a shorter period of time compared to other datasets like Imagenet~\cite{ILSVRC15} and is therefore able to serve as a proof of concept. A third reason for choosing CIFAR-10 is related with comparison purposes, since it is a popular dataset used many times in the literature.

The chosen architecture for the network is as follows:

\begin{itemize}
	\item Convolutional layer (henceforth known as $C_1$), with kernels with $5\times5$ pixels size;
	\item Normalization layer;
	\item Pooling layer, with stride 2;
	\item Convolutional layer (henceforth known as $C_2$), with kernels with $5\times5$ pixels size;
	\item Normalization layer;
	\item Pooling layer, with stride 2;
	\item Fully connected layer;
	\item Loss layer, with softmax loss.
\end{itemize}

The goal of the experiment is threefold: 1) analyze the speedup achieved using a varying number of devices; 2) quantify the influence that the number of kernels in each convolutional layer have on the speedups and 3) evaluate how the batch size impacts the speedups.

To achieve that, the number of kernels on each convolutional layer was varied, testing 4 different network architectures. The smallest tested \ac{CNN} has $50$ kernels in the first convolutional layer and $500$ on the second one. The next architectures use $150$ and $300$ kernels for the first layer and $800$ and $1000$ kernels for the second one, while the largest tested network has $500$ and $1500$ kernels on each layer, respectively.

\subsection{Experimental results}

This section presents the results obtained by applying the distribution devised previously and assesses its performance for the two case studies considered: \ac{CPU} and \ac{GPU} clusters. This section begins with an analysis of the speedup achieved using a variation in number of devices. On a second stage, the effects of batch size and number of kernels per convolutional layer are also considered for this type of distribution method. Finally, a comparison of the overall performance of the \ac{CPU} and \ac{GPU} (and combinations of both), considering the same experimental parameters is performed.

\subsubsection{Speedups using CPU-cluster}

PC1 serves as the master node for the \ac{CPU} implementation, being the reference of comparison when using a single \ac{CPU}. The rest of the devices considered, PC2, PC3 and PC4 are introduced in this order to test the introduction of more nodes for the cases with two, three, and four devices, respectively.

\begin{figure*}[ht!]
	\centering
	\begin{subfigure}[b]{0.47\textwidth}
		\includegraphics[width=\textwidth]{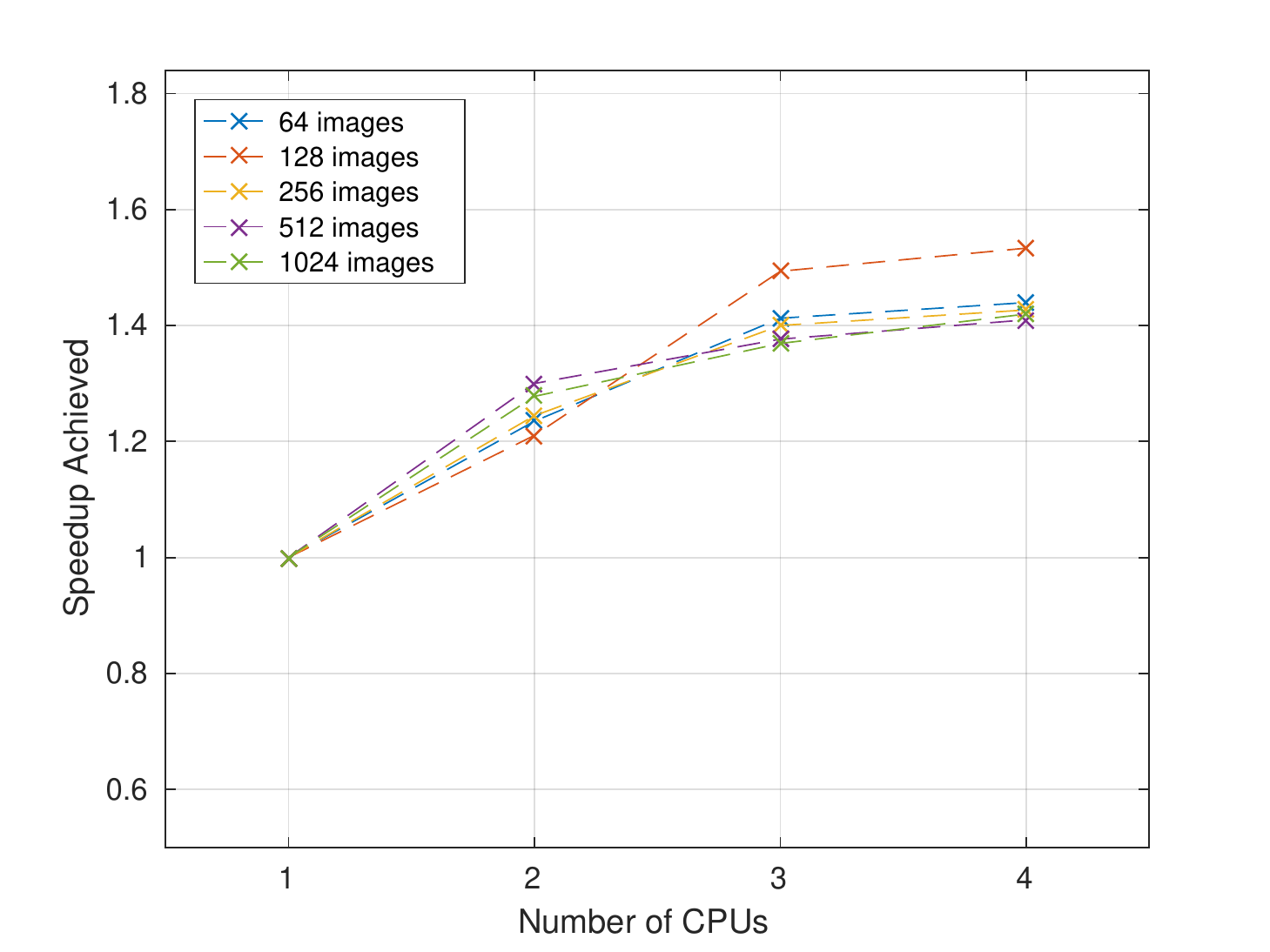}
		\caption{$C_1 = 50$ and $C_2 = 500$ kernels.}
		\label{SpeedUp_Architecture1_CPU}
	\end{subfigure}
	\quad
	\begin{subfigure}[b]{0.47\textwidth}
		\includegraphics[width=\textwidth]{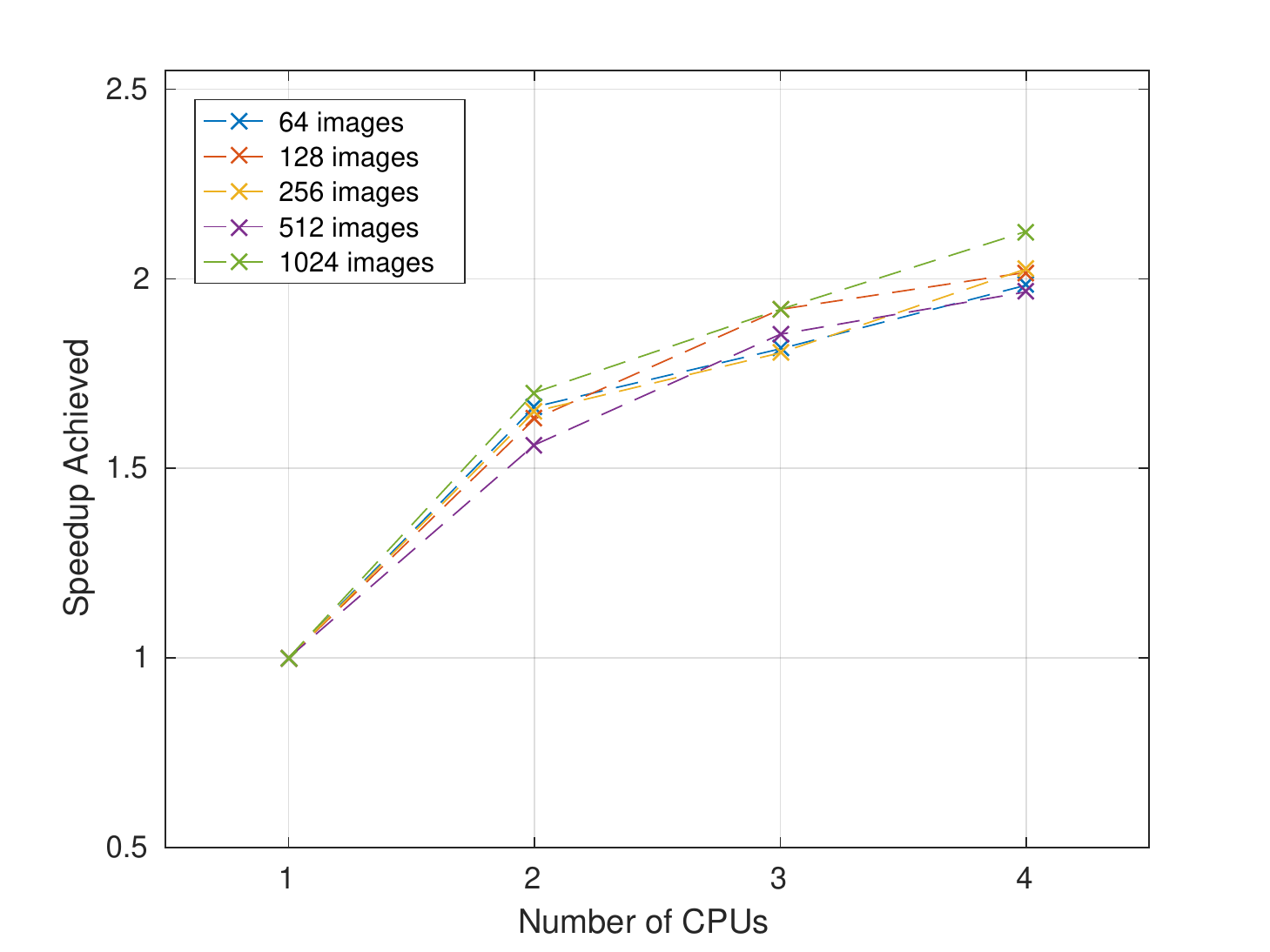}
		\caption{$C_1 = 150$ and $C_2 = 800$ kernels.}
		\label{SpeedUp_Architecture2_CPU}
	\end{subfigure}
	
	\begin{subfigure}[b]{0.47\textwidth}
		\includegraphics[width=\textwidth]{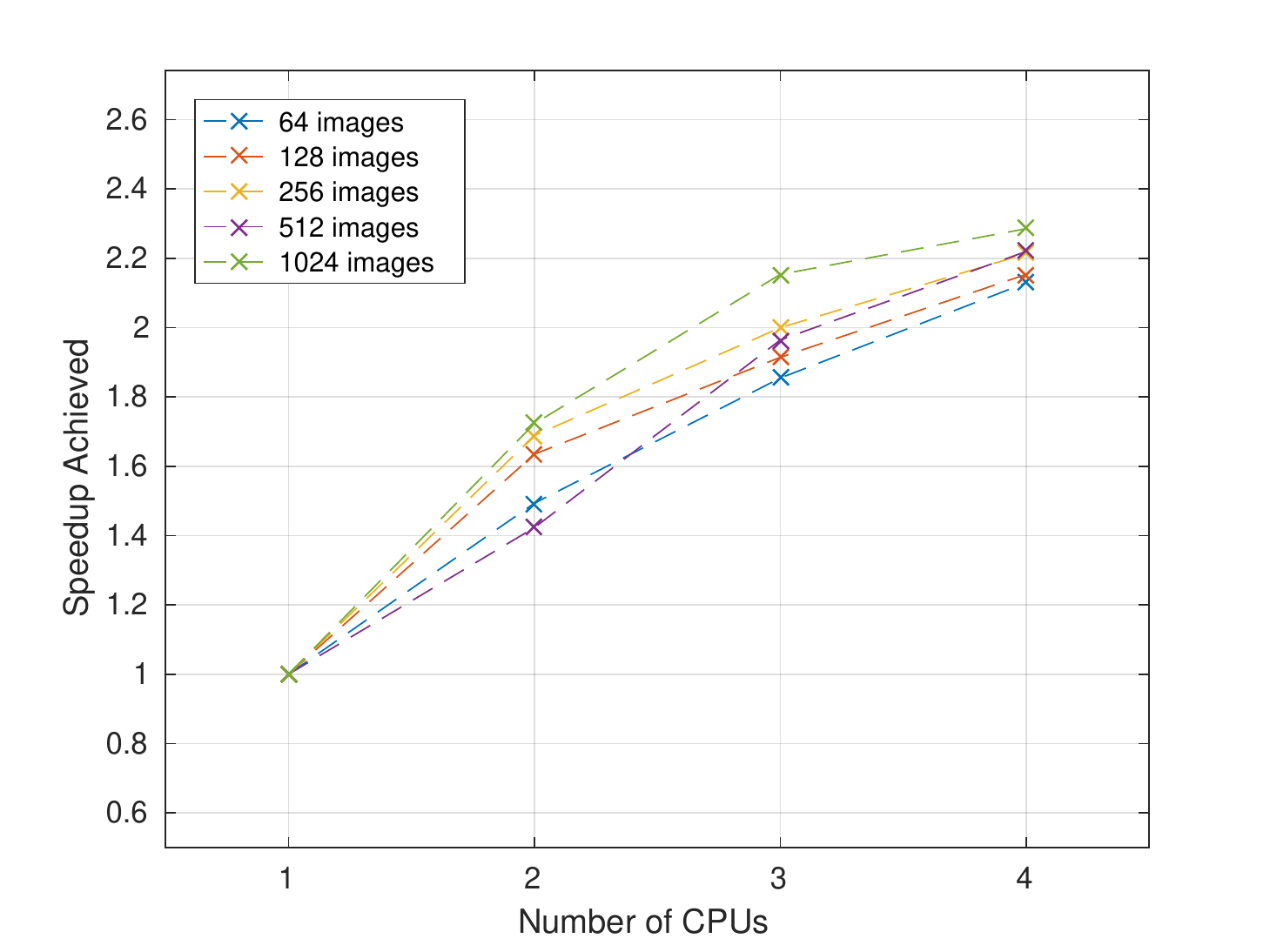}
		\caption{$C_1 = 300$ and $C_2 = 1000$ kernels.}
		\label{SpeedUp_Architecture3_CPU}
	\end{subfigure}
	\quad
	\begin{subfigure}[b]{0.47\textwidth}
		\includegraphics[width=\textwidth]{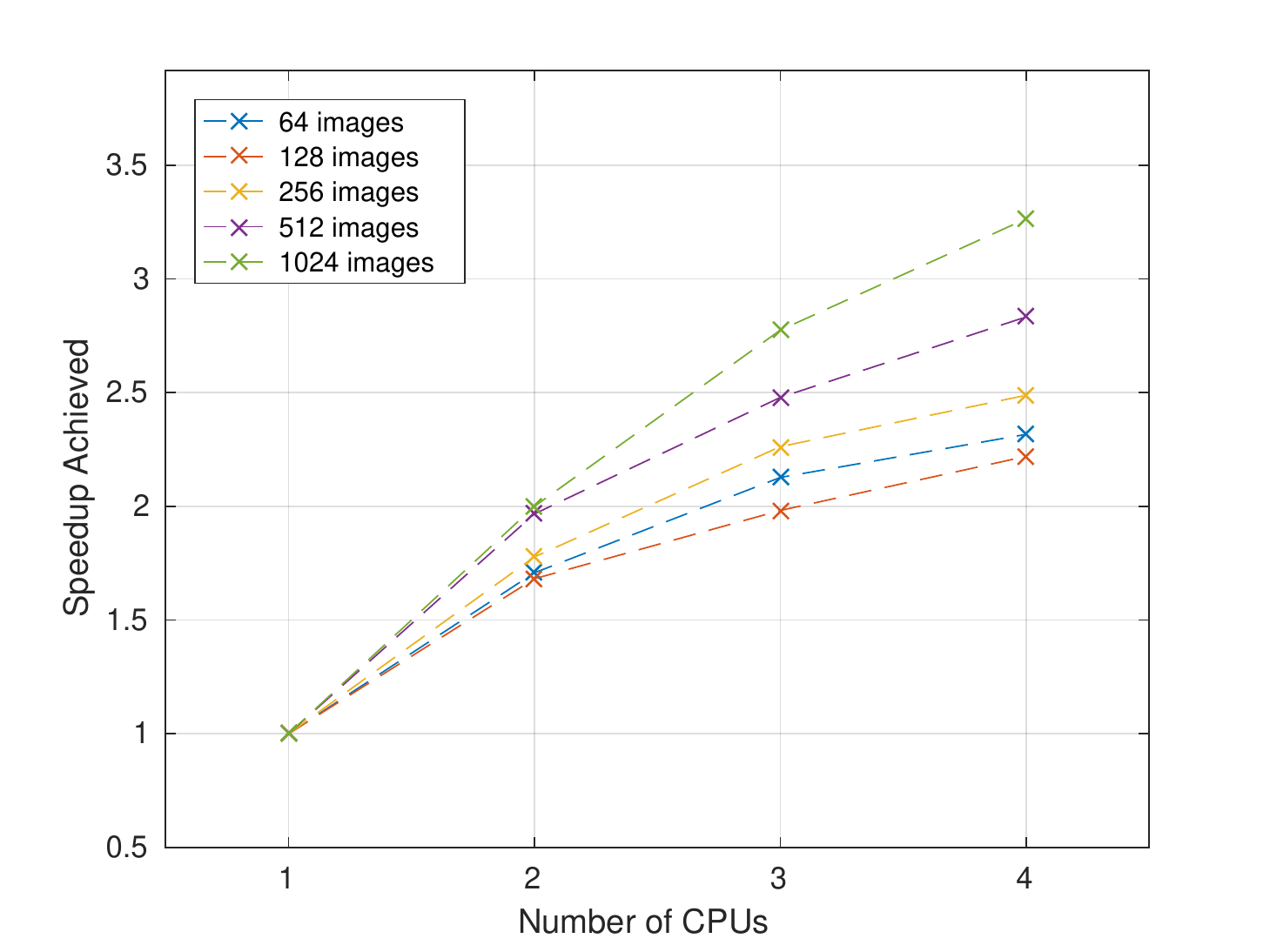}
		\caption{$C_1 = 500$ and $C_2 = 1500$ kernels.}
		\label{SpeedUp_Architecture4_CPU}
	\end{subfigure}
	\caption{Attained speedup for all batch sizes using different architectures, using a CPU cluster ranging from 1 to 4 PCs.}
	\label{SpeedUp_Architecture_CPU}
\end{figure*}

All subfigures in Figure \ref{SpeedUp_Architecture_CPU} document the results by maintaining the same network architecture and varying the batch size, thus it is possible to understand the influence of batch sizes by analyzing each subfigure individually, and to study how the number of kernels affects the attained speedup by comparing the result of each batch size across different subfigures.

By analyzing Figure \ref{SpeedUp_Architecture_CPU}, it is visible that a speedup always exists, even when considering the smallest network and batch size. Looking at the results of subfigure \ref{SpeedUp_Architecture1_CPU}, it is noticeable that the introduction of more \acp{CPU} contributes to an improvement on processing time, achieving speedups of $1.3\times$ for $2$ CPUs, $1.5\times$ for $3$ and slightly above $1.5\times$ for $4$ CPUs.
\newline

\textit{Batch Size}: A batch size influence analysis on the distribution technique performance starts by comparing each subfigure individually. For the smallest considered network, the difference in batch size does not introduce significant changes, since the speedups attained for 4 CPUs are between $1.4\times$ and $1.55\times$.

For the two next architectures, with $150$ kernels on the first convolutional layer and $800$ kernels on the second one, and $300$ kernels on the first layer and $1000$ on the second, the differences continue to be almost non existent: there is a performance gain accompanying the increase in convolutional layer size that is considerably constant with the increment of batch size. These two networks achieve speedups of $1.95\times$ to $2.1\times$ and $2.1\times$ to $2.3\times$, respectively.

However, for the largest network tested, there is a more prominent difference when training it with different batch sizes, with the speedups for $4$ CPUs ranging from $2.21\times$ to $3.28\times$. Nonetheless, the difference of speedups between different batch sizes across the other trained networks is very small.
\newline

\begin{figure*}[htpb]
	\centering
	\begin{subfigure}[b]{0.47\textwidth}
		\includegraphics[width=\textwidth]{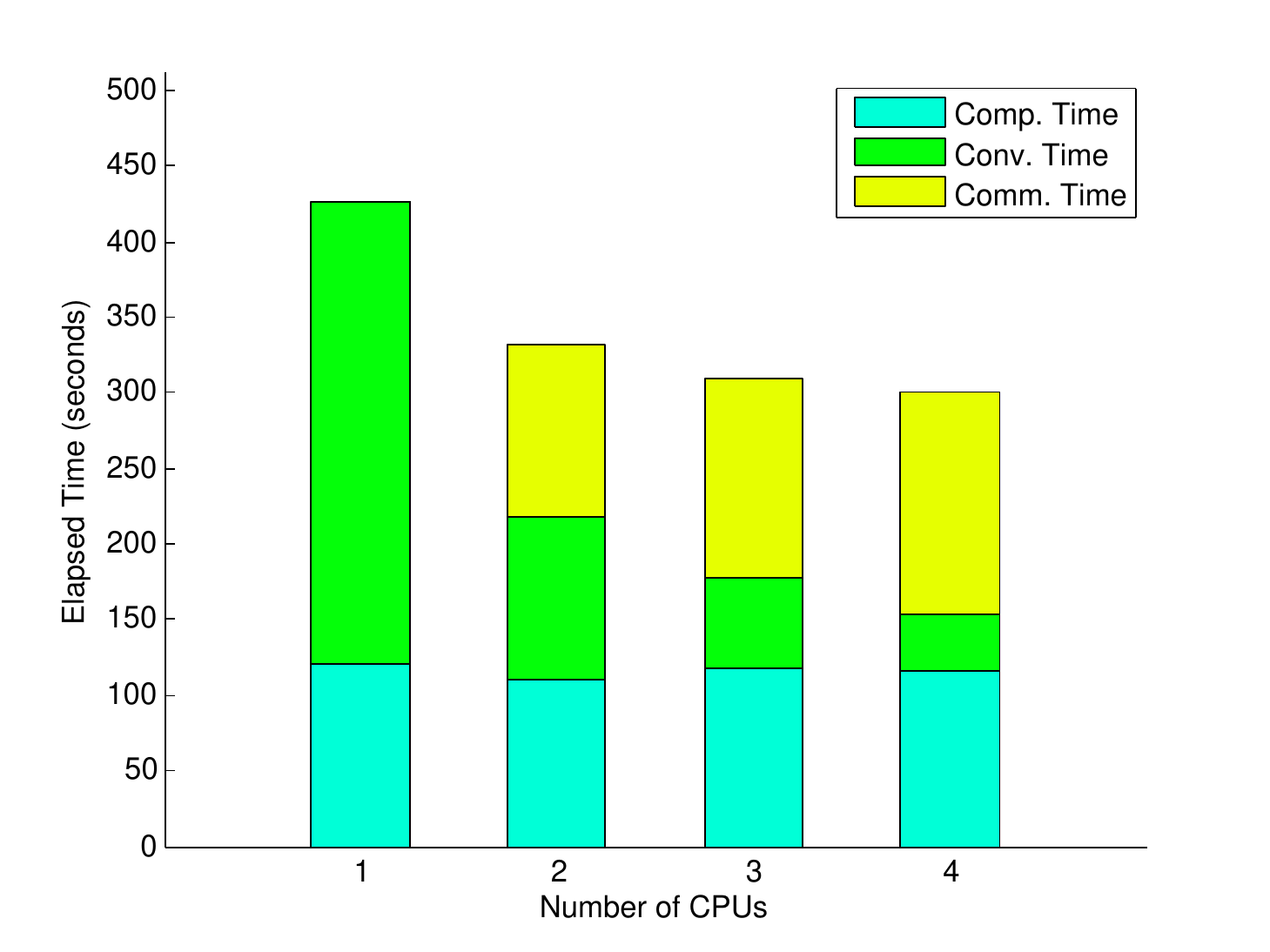}
		\caption{$C_1 = 50$ kernels and $C_2 = 500$ kernels.}
		\label{B_1024-C_1-CPU}
	\end{subfigure}
	\quad
	\begin{subfigure}[b]{0.47\textwidth}
		\includegraphics[width=\textwidth]{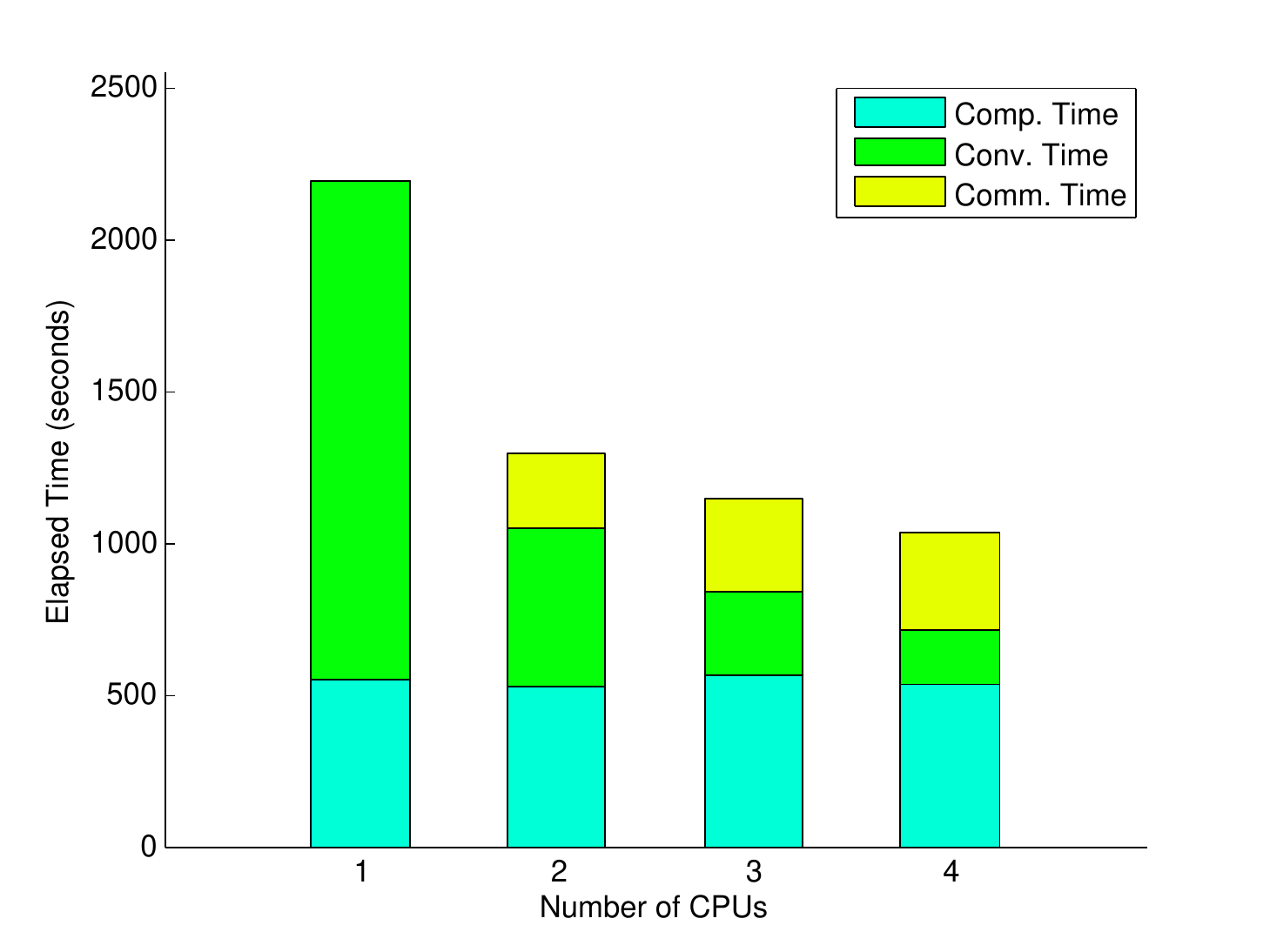}
		\caption{$C_1 = 150$ kernels and $C_2 = 800$ kernels.}
		\label{B_1024-C_2-CPU}
	\end{subfigure}
	
	\begin{subfigure}[b]{0.47\textwidth}
		\includegraphics[width=\textwidth]{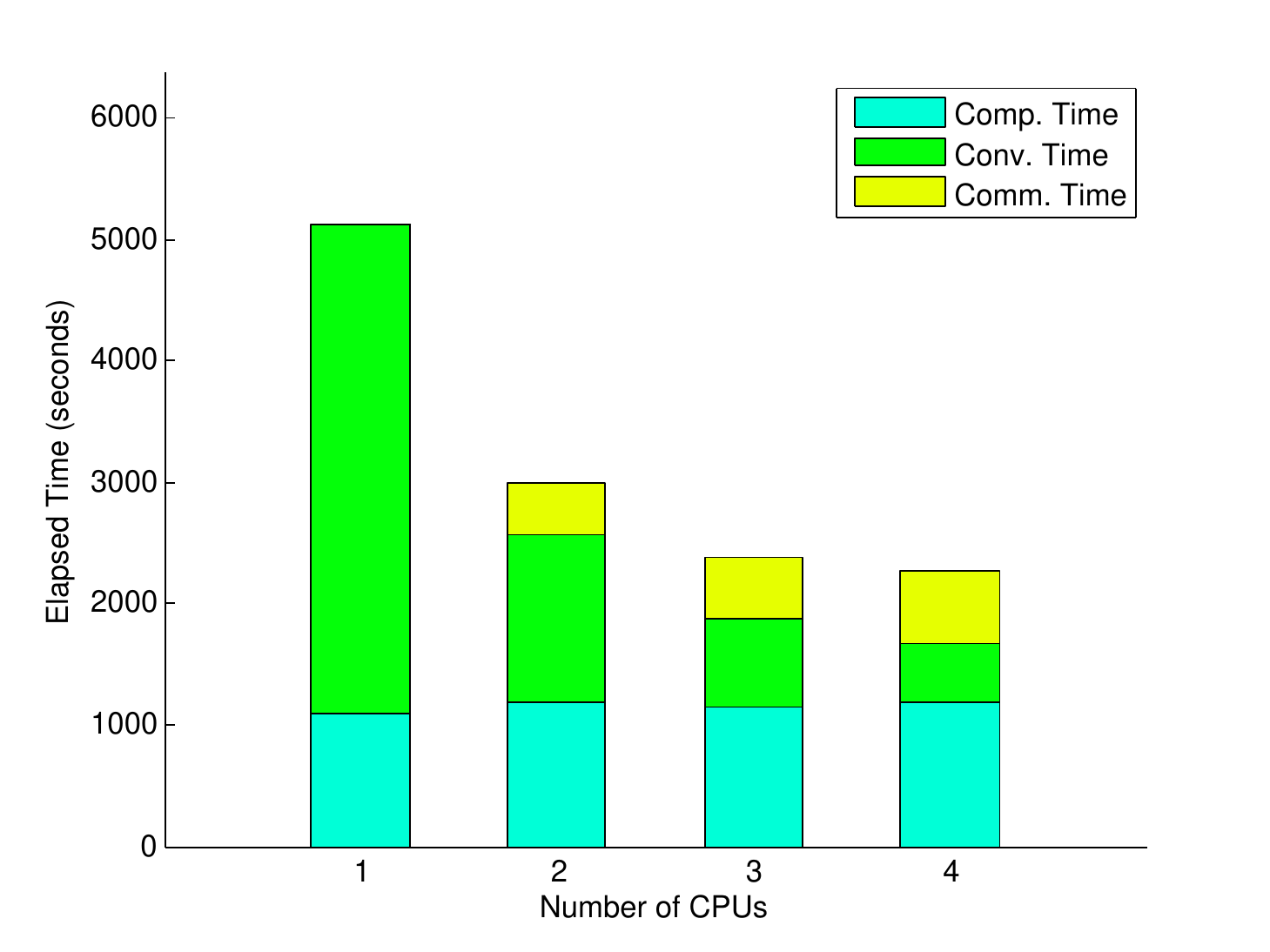}
		\caption{$C_1 = 300$ kernels and $C_2 = 1000$ kernels.}
		\label{B_1024-C_3-CPU}
	\end{subfigure}
	\quad
	\begin{subfigure}[b]{0.47\textwidth}
		\includegraphics[width=\textwidth]{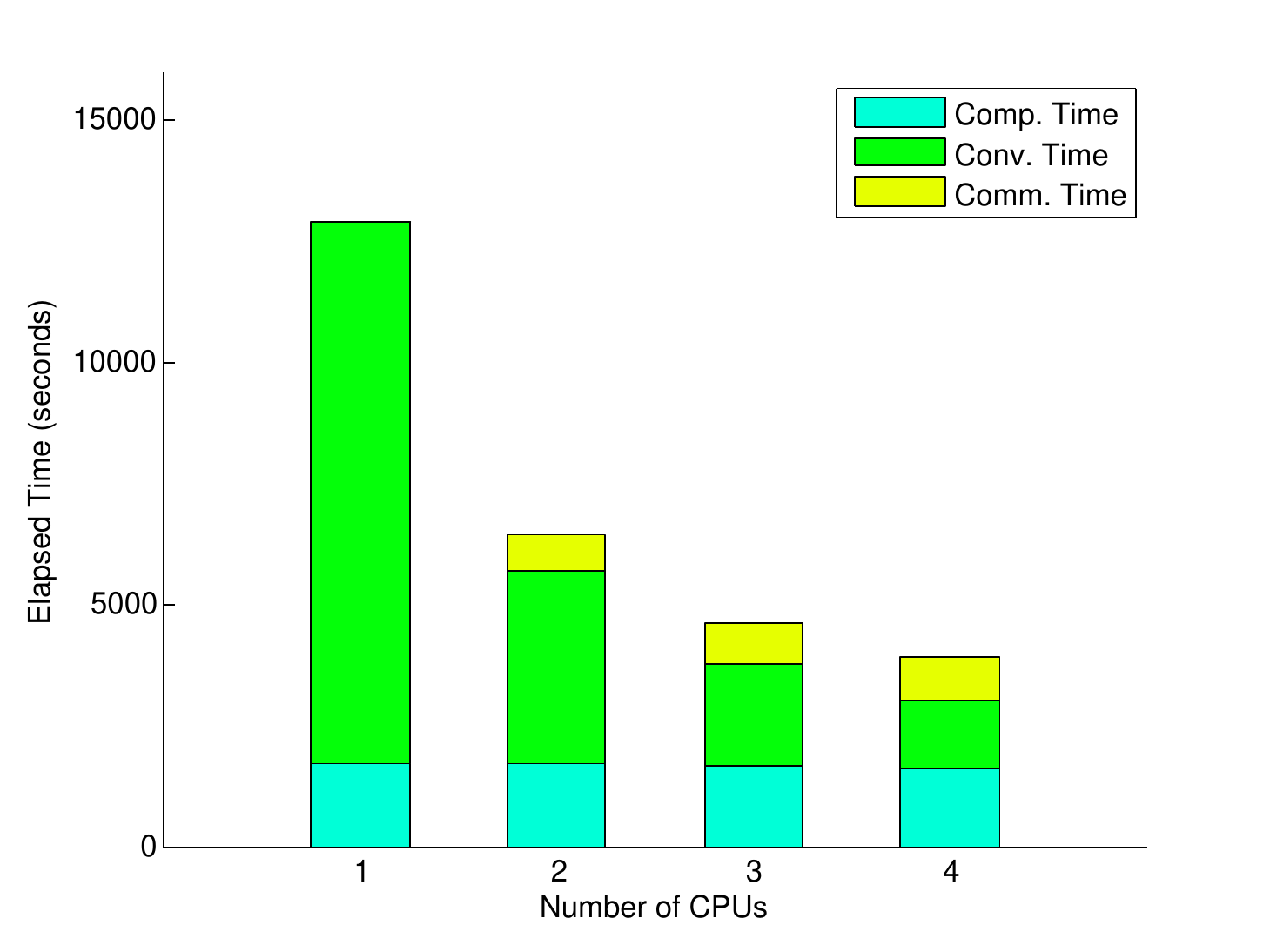}
		\caption{$C_1 = 500$ kernels and $C_2 = 1500$ kernels.}
		\label{B_1024-C_4-CPU}
	\end{subfigure}
	\caption{Elapsed time for a batch size with $1024$ images, with different architectures, using a CPU cluster ranging from $1$ to $4$ PCs.}
	\label{B_1024-CPU}
\end{figure*}

\textit{Number of Kernels}: To understand the effects that the number of kernels have on the attained speedups, it is necessary to analyze the results pertaining a batch size accross the different network architectures.

Considering the case with a batch of $64$ images, it is visible that the speedup increases from $1.45\times$ using the smallest network to  almost $2.25\times$, with the largest one.

By further analyzing the effects of the network architecture, a quick study involving the remaining batch sizes show that the increase in convolutional layers always leads to an improvement in speedup, with the worst case being with $128$ images, where the $4$ CPU case improves from 1.52x to $2.2\times$ when considering the largest network. The best case scenario comes when using a batch of $1024$ images, considering the speedups climb from $1.4\times$ to $3.25\times$ when considering $4$ CPUs training the largest network.

For the largest considered architecture, the speedups are always above $1.65\times$ with $2$ CPUs and are capable of achieving speedups greater than $2\times$ for $3$ or more CPUs. To understand how communication times differ, it is necessary to analyze how the training time is distributed. Figure \ref{B_1024-CPU} shows the elapsed time relative to only one batch of $1024$ images, since the time for the training of an entire epoch is mostly linear. The full training time is divided into $3$ parts: \textit{Comm. time} refers to the communication time between master node and slaves. \textit{Conv. time} is the time spent in convolutions by each node, or by the slowest node, as opposed to being the cumulative time spent in convolutions by all the nodes. Finally, \textit{Comp. time} is the time spent on computation of layers other than convolutional ones.

By analyzing the training time using $1$ \ac{CPU} on the different subfigures, it is visible a decrease in percentage of time dedicated to the computation of different layers, going from $25$\% with the smallest network to $13$\% when training the largest one. Considering that a more thorough analysis of the largest network using a batch with $1024$ images shows, as it can be seen in subfigure \ref{B_1024-C_4-CPU}, that the use of 2 \acp{CPU} achieves a speedup of $1.98\times$, while for $3$ and $4$ \acp{CPU} the attained speedup is $2.73\times$ and $3.28\times$, respectively. Considering that the computation of the remaining layers only occupies $13$\% of the total training time using one \ac{CPU}, the theoretical maximum speedup achievable for this particular case would be about $7.76\times$. Therefore, networks that rely more heavily on convolutions will be able to achieve better speedups.

Thus, it is possible to see that speedups are more dependent on the number of kernels than batch size, for the CPU case.

Vishnu \textit{et al.} also parallelizes a CNN training with CIFAR-10, using several CPUs connected with InfiniBand and is able to achieve a speedup of $3.01\times$ using $64$ cores, relative to the training time using $4$ cores~\cite{vishnu2016distributed}.

\subsubsection{Speedups using GPU-cluster}

In order to analyze the results obtained with the \ac{GPU} clusters, it is necessary to clarify some constraints, namely regarding the number of \acp{GPU} used in the experiments. Considering that the code was developed using MATLAB, \ac{GPU} integration is possible using CUDA, that only allows its execution on NVIDIA \acp{GPU}. Taking into account that only $3$ of the $4$ computers complete such requirement, the maximum size of the \ac{GPU} cluster is $3$ machines, which allows the comparison between \ac{CPU} and \ac{GPU} up to a certain point.

Another aspect to consider is the computational capabilities of the \acp{GPU}. As stated previously, as long as divergent operations are negligible or controlled the \ac{GPU} is much more effective than the \ac{CPU} when it comes to receiving large quantities of data and repeat the same operation, mostly sum and multiplication, very quickly, due to the large number of parallel cores available. However, for smaller amounts of data globally the \ac{GPU} handles these tasks less efficiently than the \ac{CPU} would.

Finally, as in the \ac{CPU} case, only the convolutional layers were parallelized using \ac{GPU} computing, which implies that the computation of the remaining layers is performed on the \ac{CPU}.

For this particular case, since PC1 is not a NVIDIA \ac{GPU}, the PC2 serves as the master node, also being the reference for the case of a single \ac{GPU}. PC3 and PC4 are introduced in this order to test the addition of more nodes for the cases with $2$ and $3$ devices, respectively. This notation respects the one used in Section \ref{subsec:hardware}.
\newline

\begin{figure*}[ht!]
	\centering
	\begin{subfigure}[b]{0.47\textwidth}
		\includegraphics[width=\textwidth]{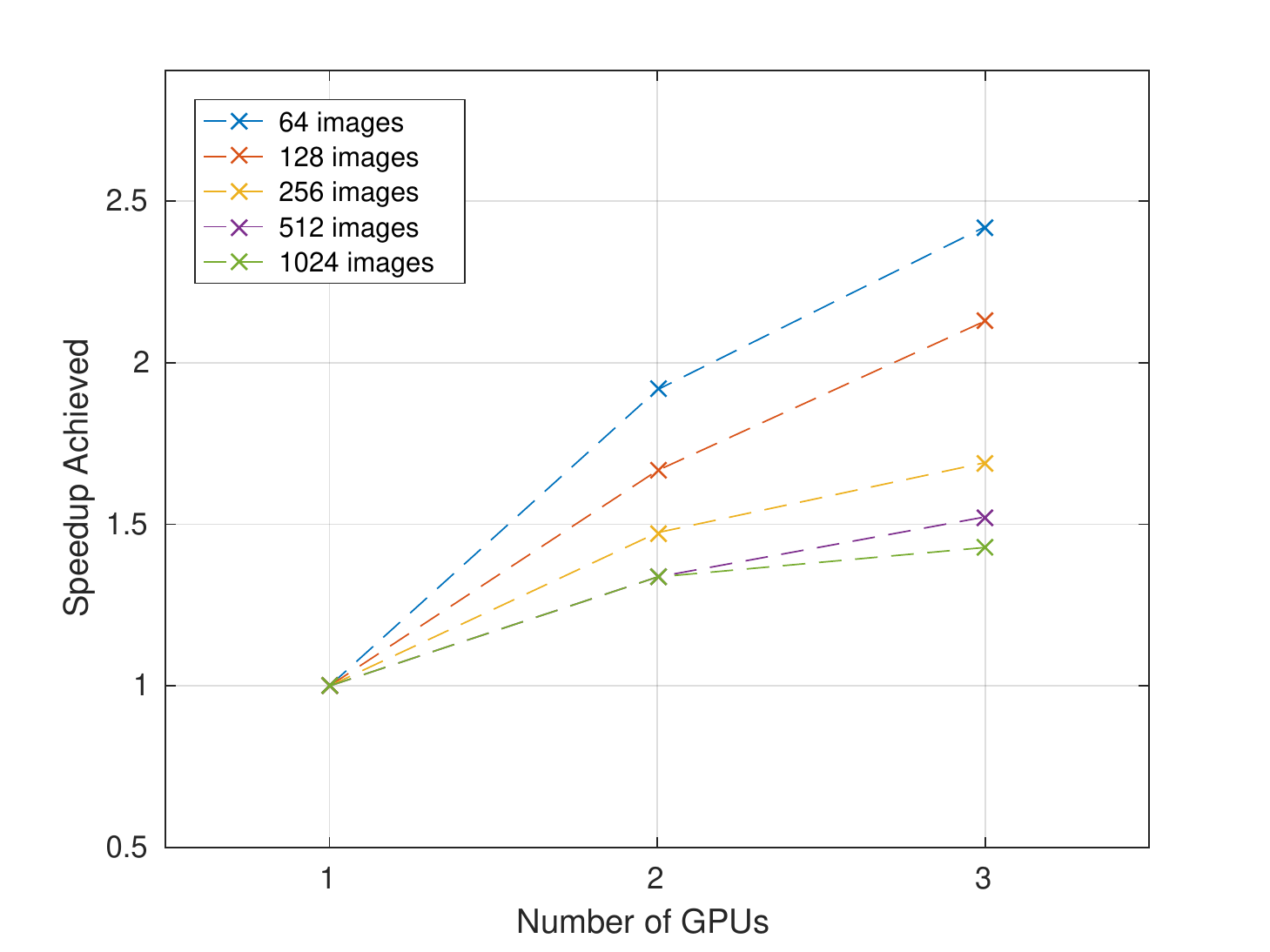}
		\caption{$C_1 = 50$ and $C_2 = 500$ kernels.}
		\label{SpeedUp_Architecture1_GPU}
	\end{subfigure}
	\quad
	\begin{subfigure}[b]{0.47\textwidth}
		\includegraphics[width=\textwidth]{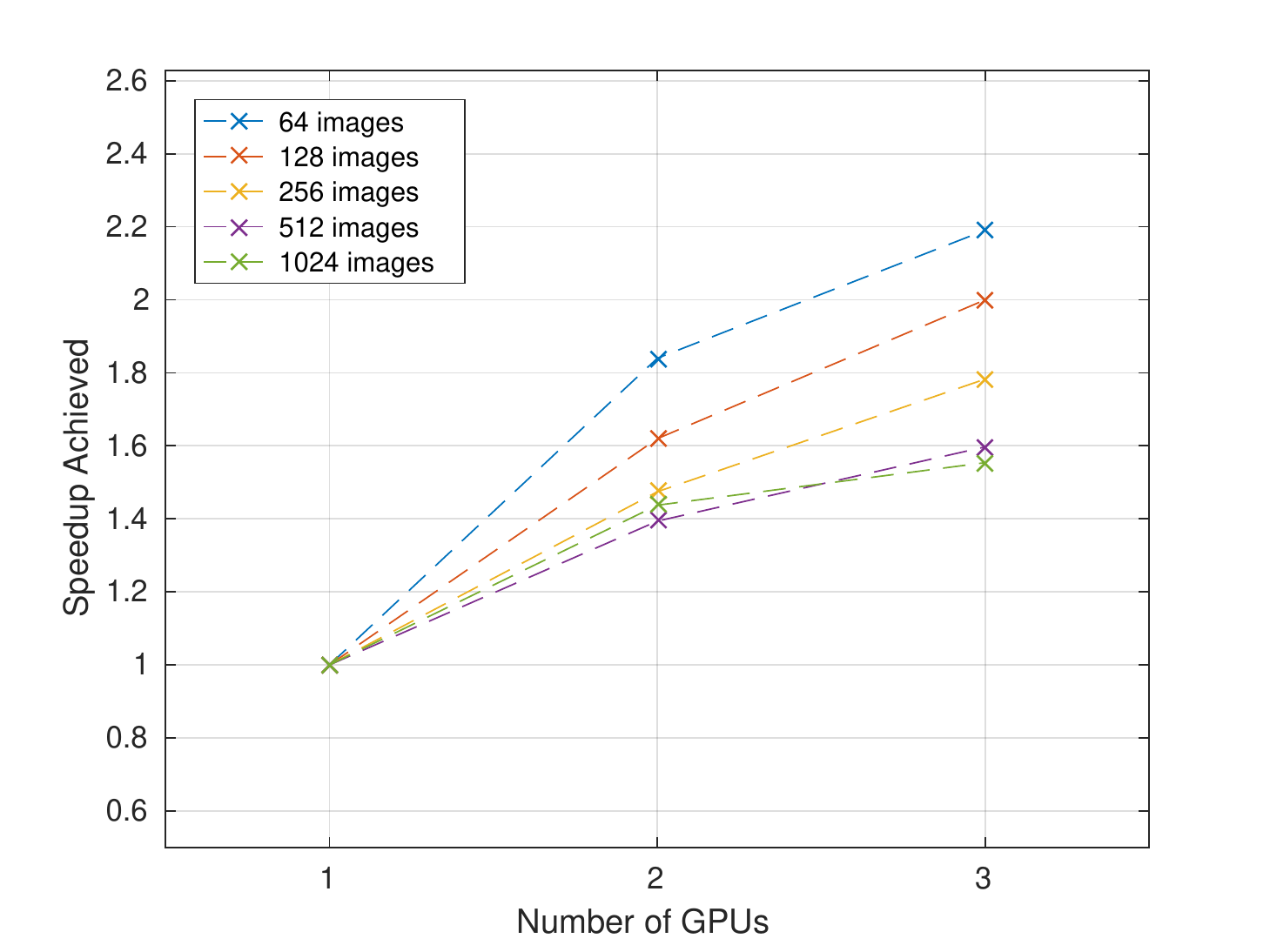}
		\caption{$C_1 = 150$ and $C_2 = 800$ kernels.}
		\label{SpeedUp_Architecture2_GPU}
	\end{subfigure}
	
	\begin{subfigure}[b]{0.47\textwidth}
		\includegraphics[width=\textwidth]{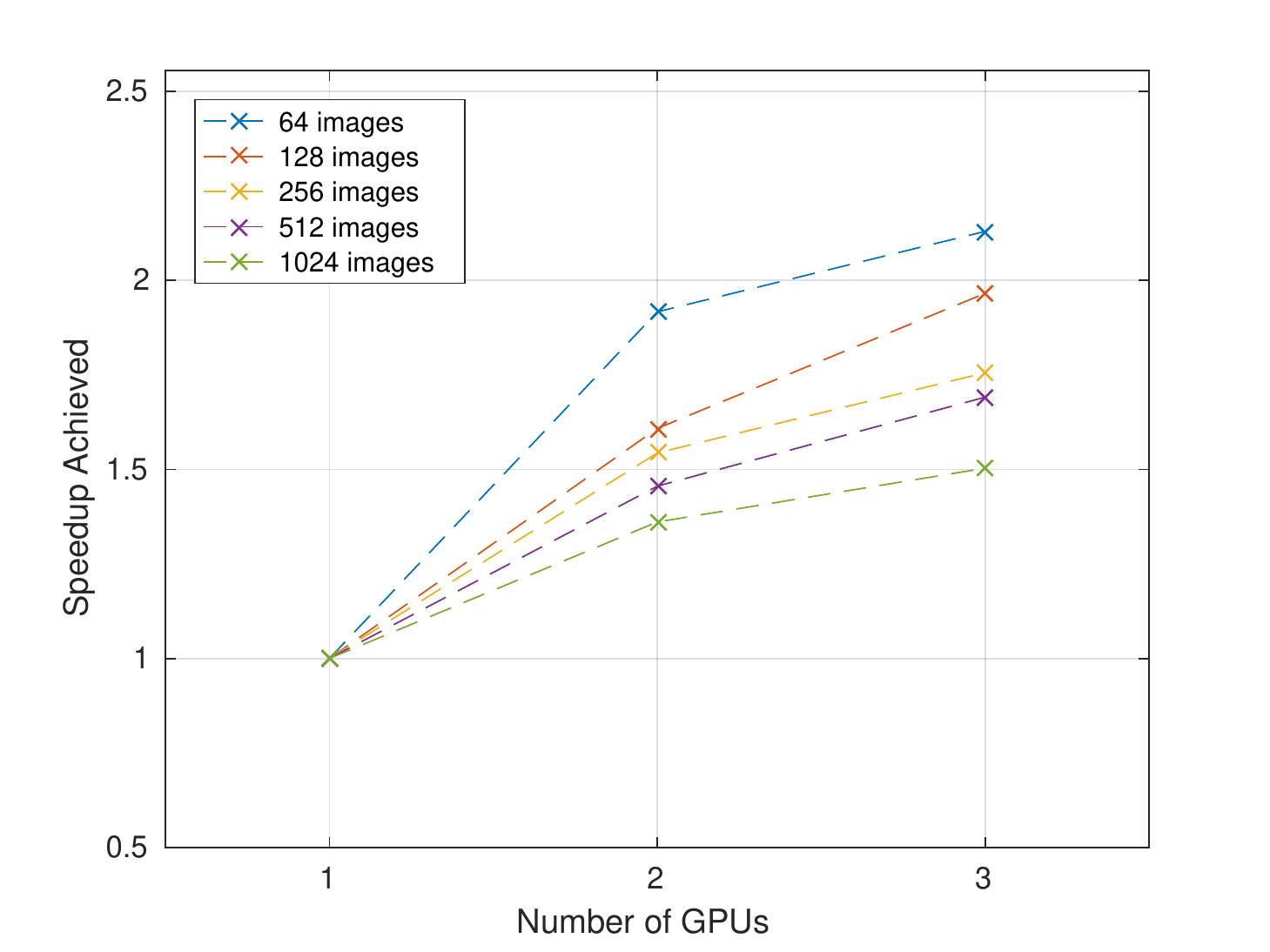}
		\caption{$C_1 = 300$ and $C_2 = 1000$ kernels.}
		\label{SpeedUp_Architecture3_GPU}
	\end{subfigure}
	\quad
	\begin{subfigure}[b]{0.47\textwidth}
		\includegraphics[width=\textwidth]{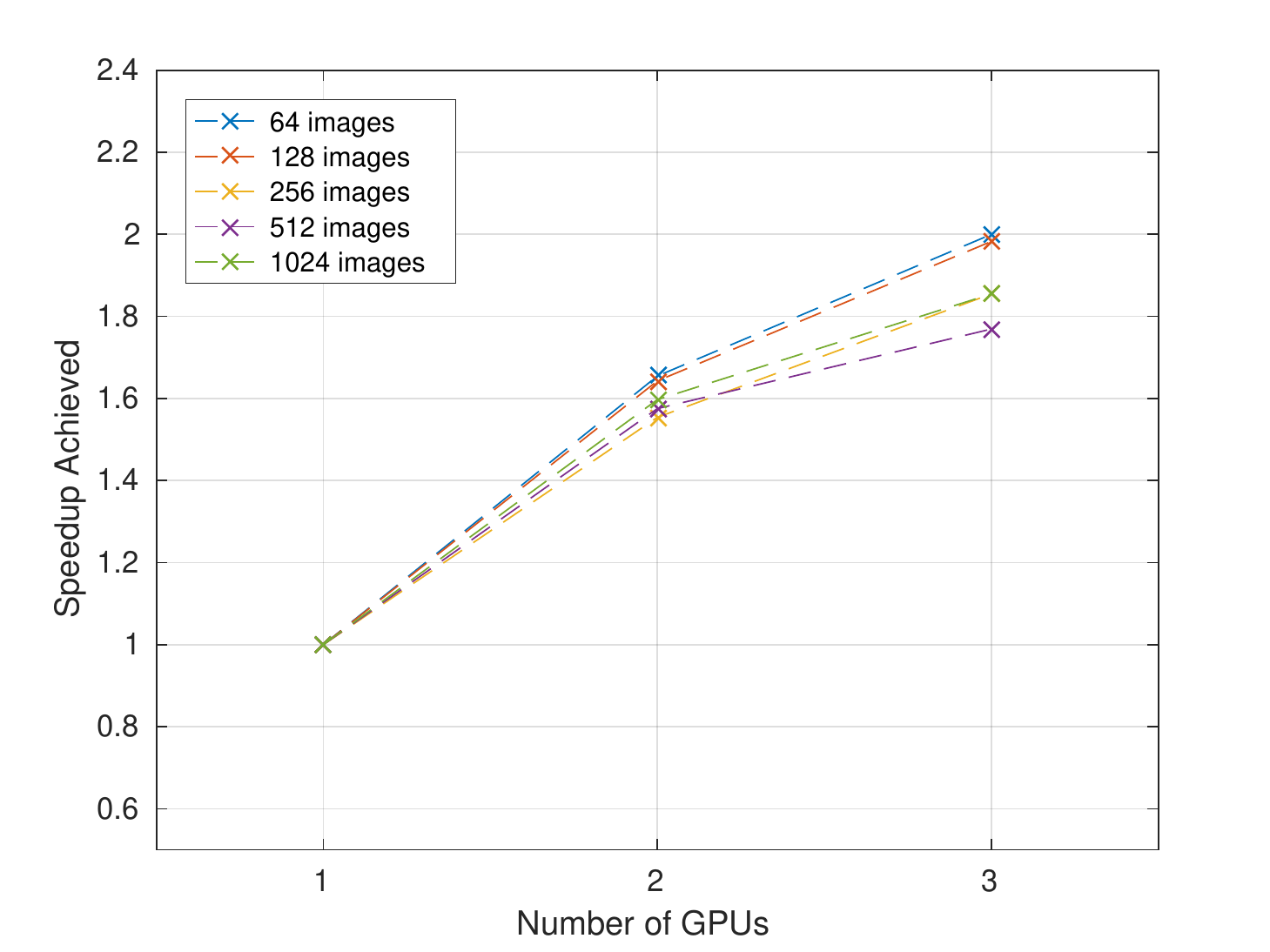}
		\caption{$C_1 = 500$ and $C_2 = 1500$ kernels.}
		\label{SpeedUp_Architecture4_GPU}
	\end{subfigure}
	\caption{Attained speedup for all batch sizes using different architectures, using a GPU cluster ranging from $1$ to $3$ PCs.}
	\label{SpeedUp_Architecture_GPU}
\end{figure*}

\textit{Batch Size}: The analysis of the influence of batch sizes on the distribution technique performance is performed by comparing each subfigure individually from Figure \ref{SpeedUp_Architecture_GPU}, like in the \ac{CPU} case. Considering the smallest trained network, there is a considerable difference between the distinct batch sizes, with speedups for $3$ GPUs ranging from $1.45\times$ to $2.45\times$.

This is a trend that continues with the next two architectures, where speedups range from $1.5\times$ to $2.2\times$ on both cases, using $3$ GPUs for the training.

However, for the largest trained network, the range of speedups is much smaller, when analyzing the different batch sizes, fluctuating between $1.75\times$ and $2\times$.
\newline

\begin{figure*}[htpb]
	\centering
	\begin{subfigure}[b]{0.47\textwidth}
		\includegraphics[width=\textwidth]{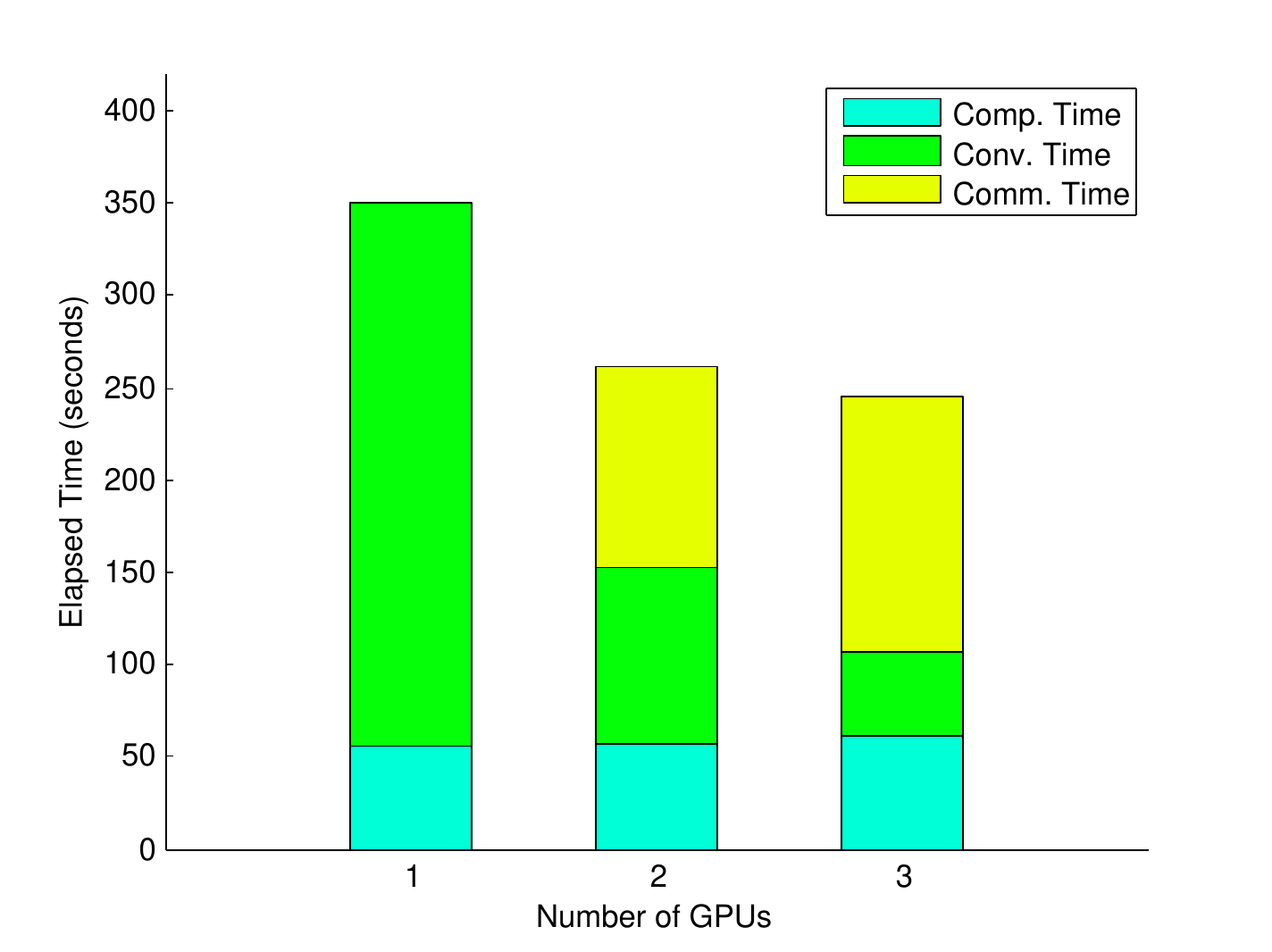}
		\caption{$C_1 = 50$ kernels and $C_2 = 500$ kernels.}
		\label{B_1024-C_1-GPU}
	\end{subfigure}
	\quad
	\begin{subfigure}[b]{0.47\textwidth}
		\includegraphics[width=\textwidth]{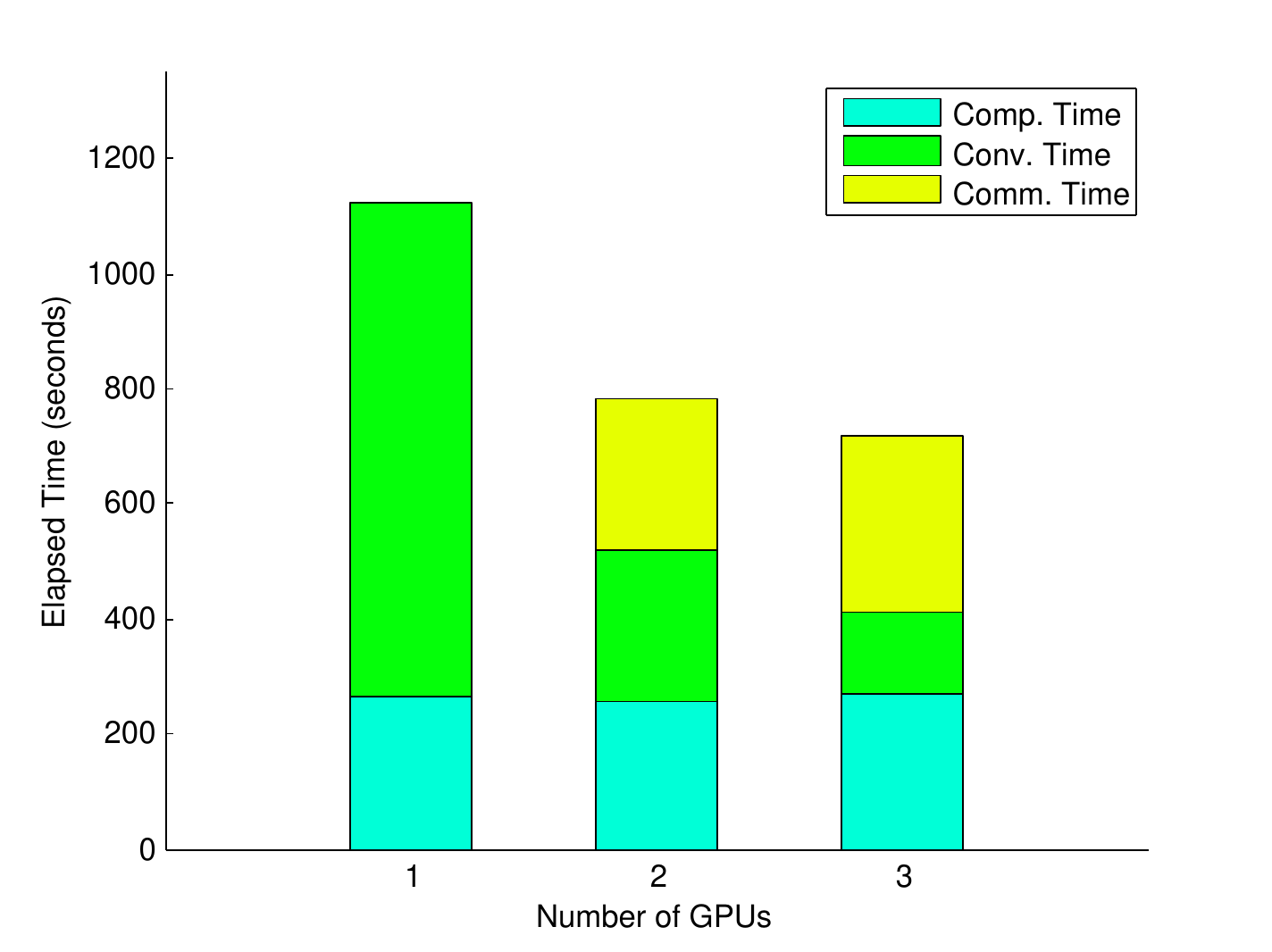}
		\caption{$C_1 = 150$ kernels and $C_2 = 800$ kernels.}
		\label{B_1024-C_2-GPU}
	\end{subfigure}
	
	\begin{subfigure}[b]{0.47\textwidth}
		\includegraphics[width=\textwidth]{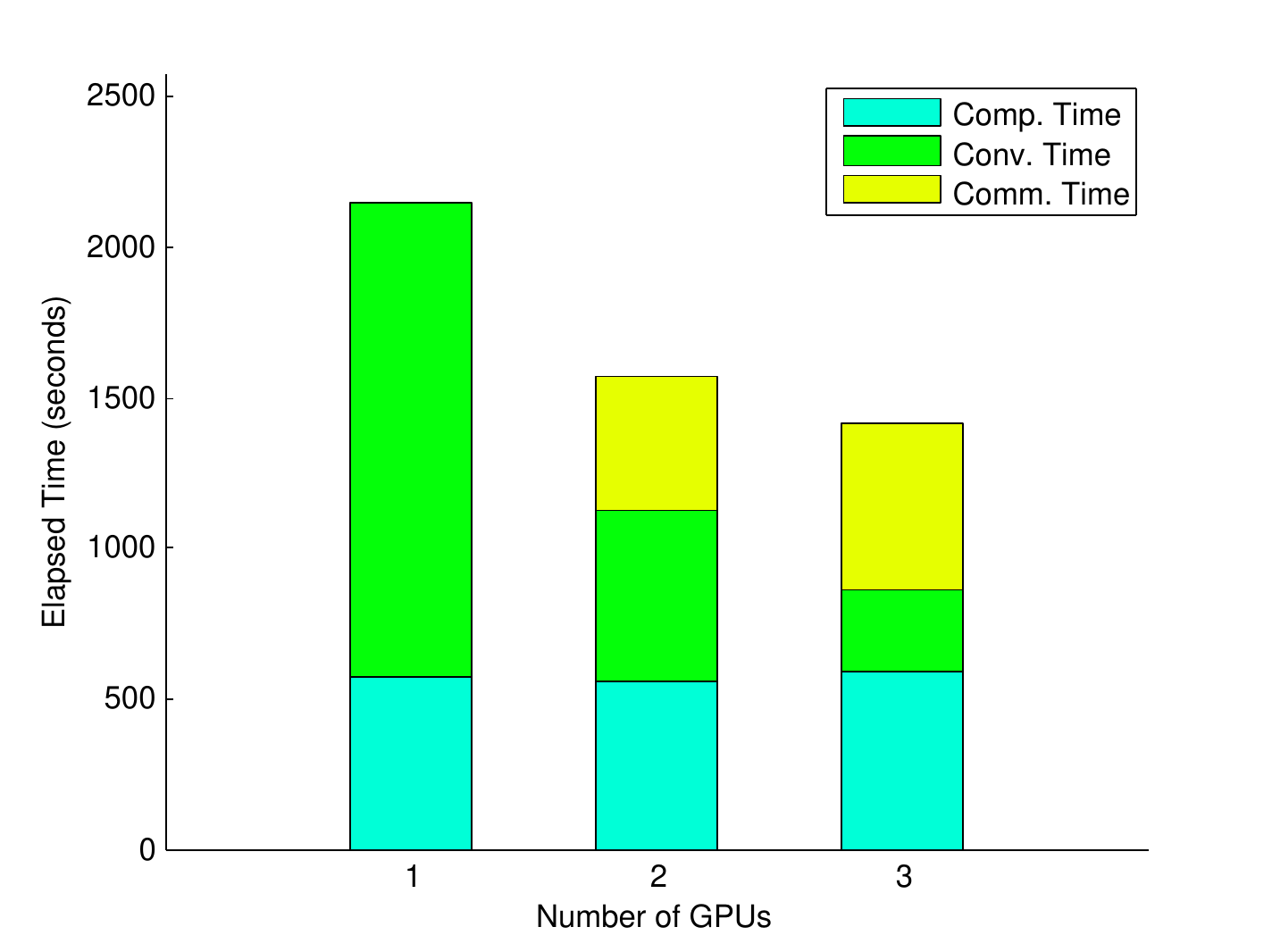}
		\caption{$C_1 = 300$ kernels and $C_2 = 1000$ kernels.}
		\label{B_1024-C_3-GPU}
	\end{subfigure}
	\quad
	\begin{subfigure}[b]{0.47\textwidth}
		\includegraphics[width=\textwidth]{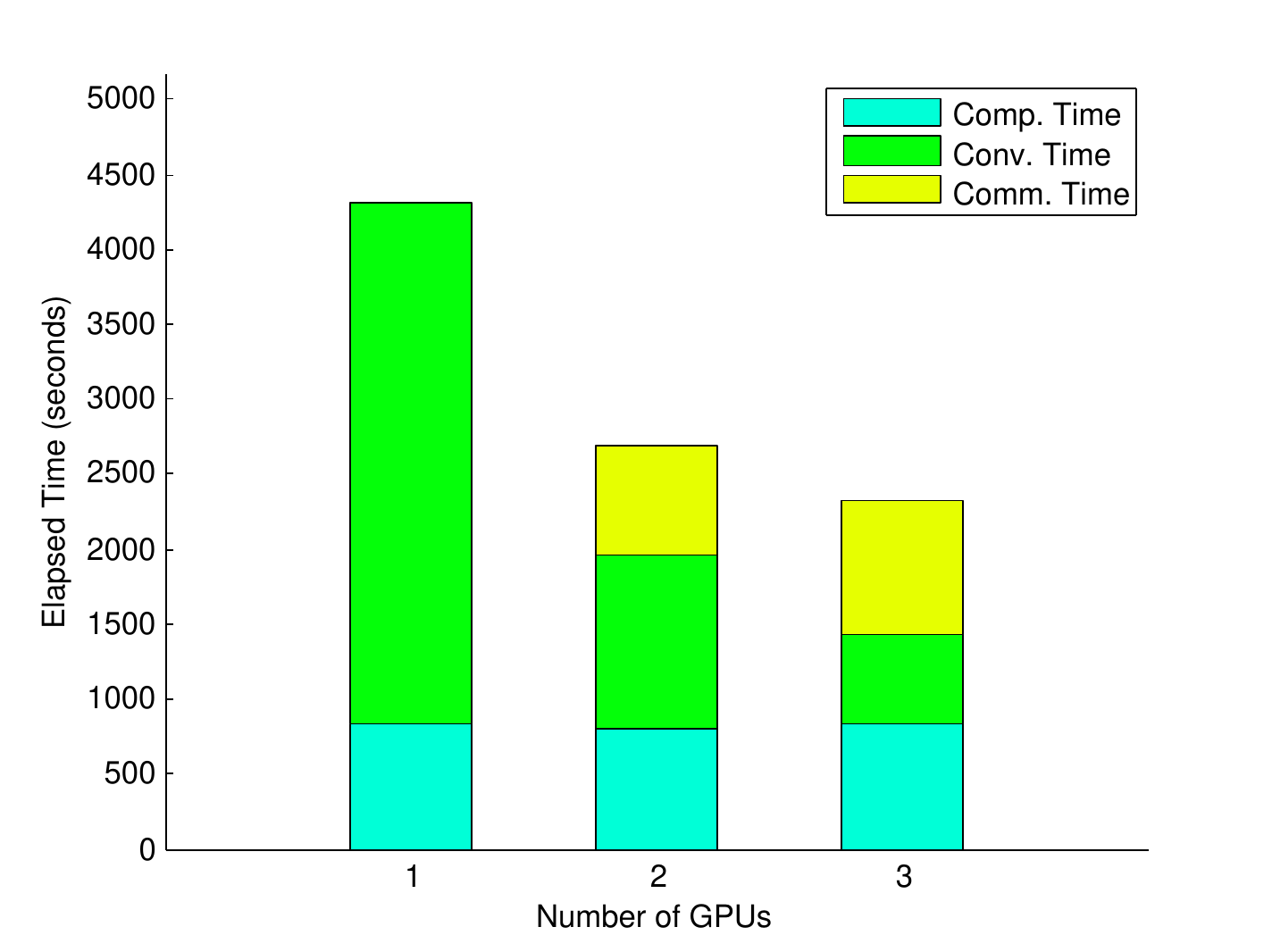}
		\caption{$C_1 = 500$ kernels and $C_2 = 1500$ kernels.}
		\label{B_1024-C_4-GPU}
	\end{subfigure}
	\caption{Elapsed time for a batch size with 1024 images, with different architectures, using a GPU cluster ranging from $1$ to $3$ PCs.}
	\label{B_1024-GPU}
\end{figure*}

\textit{Number of Kernels}: To understand the effects that the number of kernels have on the attained speedups, it is necessary to analyze the results pertaining a batch size across the different network architectures, as in the CPU case.

Considering the case with a batch of $64$ images, it is visible that the speedup decreases from $2.45\times$ using the smallest network to $2\times$, with the largest one.

This trend continues with a batch size of $128$, with speedups decreasing from $2.15\times$ to $1.95\times$, but change with the remaining batches, with speedups increasing with larger networks. To understand the differences between the CPU and GPU cases, is is necessary to analyze how the training time is distributed. Figure \ref{B_1024-GPU} shows the elapsed time relative to only one batch of $1024$ images, since the time for the training of an entire epoch is mostly linear. The full training period is divided into 3 parts: \textit{Comm. time} refers to the communication time between master node an the slaves. \textit{Conv. time} is the time spent in convolutions by each node, or by the slowest node, as opposed to being the cumulative time spent in convolutions by all nodes. Finally, \textit{Comp. time} is the time spent on computation of layers other the than convolutional ones.

As Figure \ref{B_1024-GPU} shows, an increase of kernels in the \ac{GPU} case makes almost no difference concerning communication time and speedup, and this is also visible in the rest of the tested architectures, trained using batches of $1024$ images. All the architectures tested  with this batch size show an attained speedup between $1.45\times$ and $1.80\times$ for $2$ \acp{GPU} and between $1.45\times$ and $2\times$ using $3$ \acp{GPU}, with the ratio between communication, convolution and computation time being virtually the same on the $3$ considered experiments, with the communication time rising from $19$\% with $2$ \acp{GPU} to $30$\% when using all $3$ \acp{GPU}.

The major difference between the \ac{CPU} and \ac{GPU} results is that while using the \acp{CPU}, the computation time was the major bottleneck on that experiment. However, in the \ac{GPU} case, the communication and computation time share about the same percentage of full training time, when using $3$ \acp{GPU}, which is explained by the fact that the \ac{GPU} is able to accelerate the convolutional phase.

Using the code provided by TensorFlow to train a CNN with CIFAR-10 using multiple GPUs on the same machine, it is possible to reduce the step time from $0.35-0.60$ seconds per batch with $1$ GPU to $0.13-0.20$ seconds with $2$ GPUs. However, the addition of more GPUs does not correlate to better speedups, since $3$ GPUs are able to reduce the step time to only $0.13-0.18$ seconds and $4$ GPUs still take $0.10$ seconds per batch~\cite{multigpu}.

\subsubsection{Comparison between CPU and GPU}

The following two tables show the best attained speedups for each network architecture and a given number of devices, for both \ac{CPU} (in Table \ref{tab:CPUspeedup}) and \ac{GPU} (in Table \ref{tab:GPUspeedup}). It should be noted that, for each case, speedup is obtained by comparing execution time against a single device of the same type (i.e. CPU or GPU):

\begin{table}[!h]
	\centering
	\begin{tabular}{c|ccc}
		\backslashbox{Network}{\# of CPUs} & 2 & 3 & 4 \\\hline
		50:500 & 1.40x & 1.51x & 1.56x \\
		150:800 & 1.68x & 1.93x & 2.10x \\
		300:1000 & 1.69x & 2.15x & 2.33x \\
		500:1500 & 1.98x & 2.74x & 3.28x \\
	\end{tabular}
	\caption{Best speedups achieved by network architecture and number of \acp{CPU} used.}
	\label{tab:CPUspeedup}
\end{table}

\begin{table}[!h]
	\centering
	\begin{tabular}{c|cc}
		\backslashbox{Network}{\# of GPUs} & 2 & 3 \\\hline
		50:500 & 1.96x & 2.45x \\
		150:800 & 1.89x & 2.23x \\
		300:1000 & 1.78x & 2.09x \\
		500:1500 & 1.66x & 2.00x \\
	\end{tabular}
	\caption{Best speedups achieved by network architecture and number of \acp{GPU} used.}
	\label{tab:GPUspeedup}
\end{table}

As Table \ref{tab:CPUspeedup} shows, the difference between speedups using multiple \acp{CPU}, for a given architecture, increases with the growing convolutional layers. The speedup improvement using $2$ \acp{CPU} is particularly small, although it reaches $1.98\times$ on the largest tested network. However, this tendency fades with the increase in \acp{CPU}. By training the network with $3$ \acp{CPU}, the speedup is $1.93\times$ for the second smallest network, reaching $2.74\times$ for the largest architecture. Using $4$ \acp{CPU} gives a considerable gain in speedup, particularly for the network with $300$ kernels on the first convolutional layer and $1000$ kernels on the second one, and the largest trained network. This is explained with the increase in communication time due to sending dozens more kernels to other nodes, that are only a couple of KBs, being counterbalanced by convolutions' parallelization.

However, for the \ac{GPU} implementation case, in Table \ref{tab:GPUspeedup}, the values of the speedups diminish with the enlargement of the convolutional layers. This happens because although the GPU is being used more efficiently with larger networks, the addition of more devices incurs in larger communication times, due to the need of sending a substantially higher number of kernels to the other devices. Thus, for larger networks, the attained speedup is significantly less than for smaller ones.

\subsubsection{Scalability}

As with other methods for distributed learning, speedups may only exist when using up to a certain number of nodes. For a more detailed study on scalability, it is necessary to analyze some details regarding the experiments conducted. First, the amount of data transmitted between master and slave nodes on the convolutional layers. This depends only on the number of convolutional layers and the size of their inputs, including width, height and number of input channels, size and number of kernels and the batch size. Taking this information into consideration, the number of elements $upload$ that are necessary to exchange between master and slave nodes can be described as:

\begin{multline}
upload = \sum_{i = 1}^{layers} in_i^2 \times inCh_i \times batch +\\
k_i^2 \times numK_i \times inCh_i + out_i^2 \times numK_i \times batch ,
\end{multline}
where $layers$ refers to the number of convolutional layers that need to be distributed, $in$ is the convolutional layer's input width or height, considering a square image, like this particular case, $inCh$ represents the input channels, $k$ is the kernel size, $numK$ represents the number of kernels for each convolutional layer, $out$ refers to the output's size and $batch$ is the batch size. All values transmitted are of type \textit{double}. The next detail to consider is the velocity at which the data is transmitted across nodes. A quick study of the several results achieved shows that the bandwidth is approximately constant, averaging at $5$ Mbps. Another aspect to consider is the number of kernels that should be passed to each worker, which is explored more in detail in Section \ref{subsec:hybrid}.

\begin{figure}
	\centering
	\begin{subfigure}[b]{0.47\textwidth}
		\includegraphics[width=\textwidth]{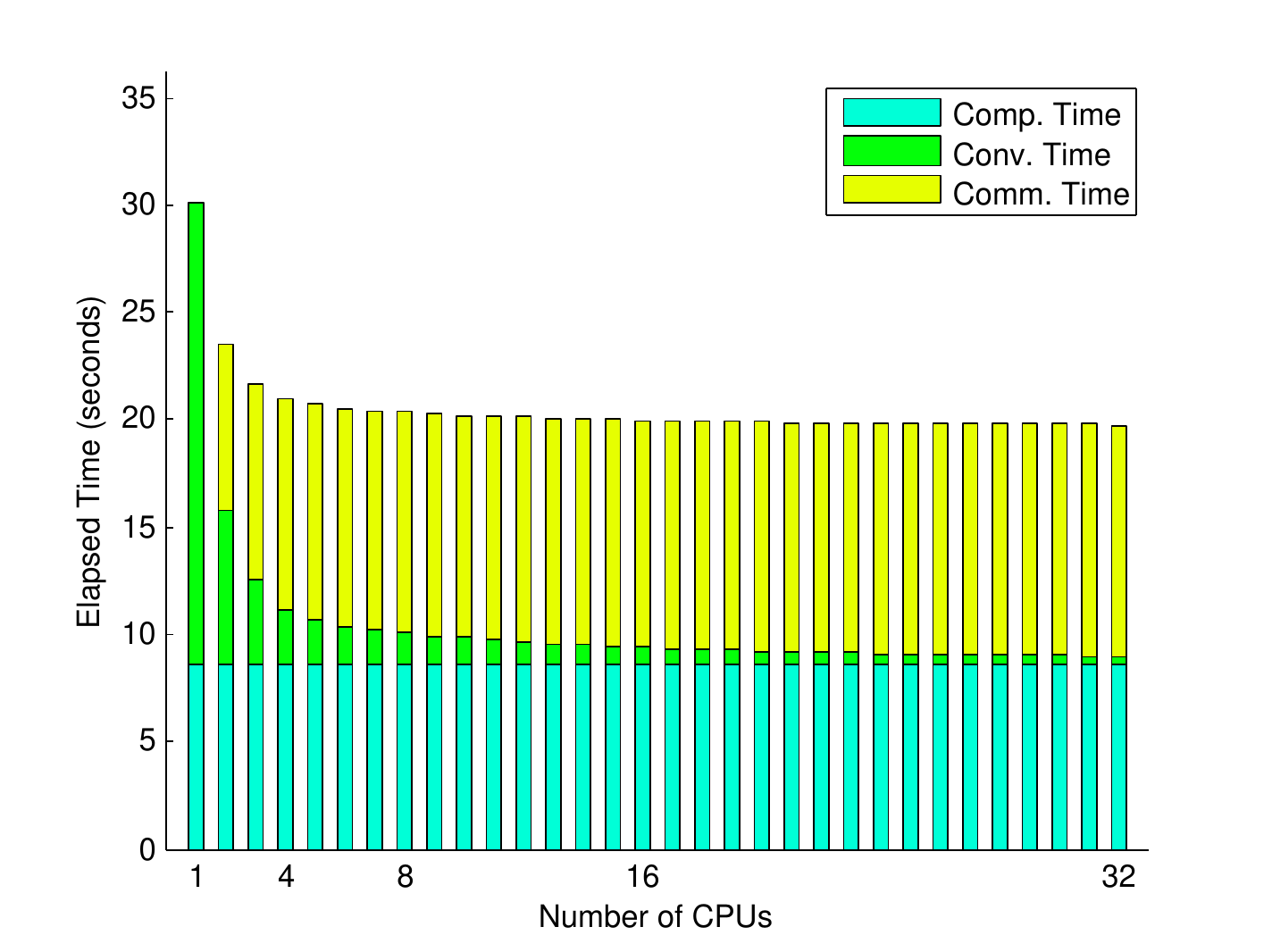}
		\caption{$C_1 = 50$ kernels and $C_2 = 500$ kernels.}
		\label{fig:Net_5-50-500-PC_32}
	\end{subfigure}
	\quad
	\begin{subfigure}[b]{0.47\textwidth}
		\includegraphics[width=\textwidth]{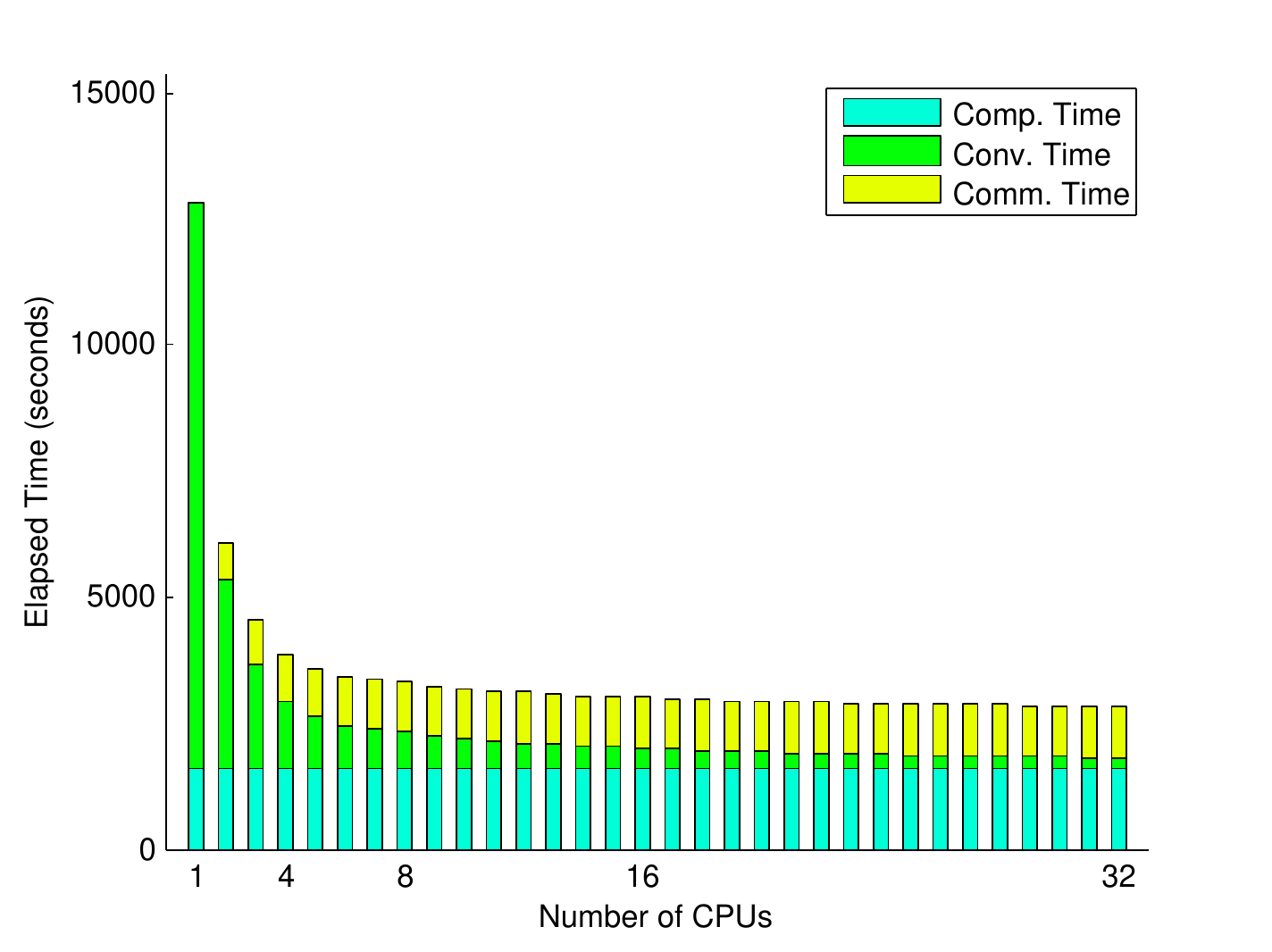}
		\caption{$C_1 = 500$ kernels and $C_2 = 1500$.}
		\label{fig:Net_5-500-1500-PC_32}
	\end{subfigure}
	\caption{Elapsed time for the smallest network, using a batch with 64 images, and the largest network, with a batch size of $1024$ images, using a CPU cluster ranging from $1$ to $32$ PCs.}
	\label{fig:Net_5}
\end{figure}

By understanding these details, it is possible to accurately predict new communication times when more nodes are added, as well as convolution times and therefore the total processing time. Three different cases where considered. The first two pertain to the \ac{CPU} case (depicted in Figure \ref{fig:Net_5}), where the processing time for the smallest and largest networks were simulated adding $32$ \acp{CPU} nodes. For these simulations, the \acp{CPU} were considered to have computational capabilities similar to the devices used in the experiment (shown in Table~2), being assigned random performance values with Gaussian distribution, varying between worst and best case scenario as depicted in Table~2 for each \acp{CPU} used. These results are shown in Figure \ref{fig:Net_5}.

As results show, the method is scalable without incurring in performance loss, despite becoming irrelevant the introduction of more nodes after a certain value. Both the case of the smallest network as the largest one benefit little from the addition of more nodes from $4$ \acp{CPU}, and there is a stabilization in speedup after $8$ nodes. This occurs because the inclusion of more \acp{CPU} leads to a slight increase in information to be sent by the master node, that is counterbalanced by the decrease in time obtained by the parallelization. It is also possible to notice that while using $1$ \ac{CPU}, the convolution time is the bottleneck. But when using several \acp{CPU}, this situation is reversed and the communication and computation times become the bottlenecks. The former can be solved with faster data transmission, but the latter can only be fixed with parallelization.

\begin{figure}[h!]
	\begin{center}
		\includegraphics[width=\columnwidth]{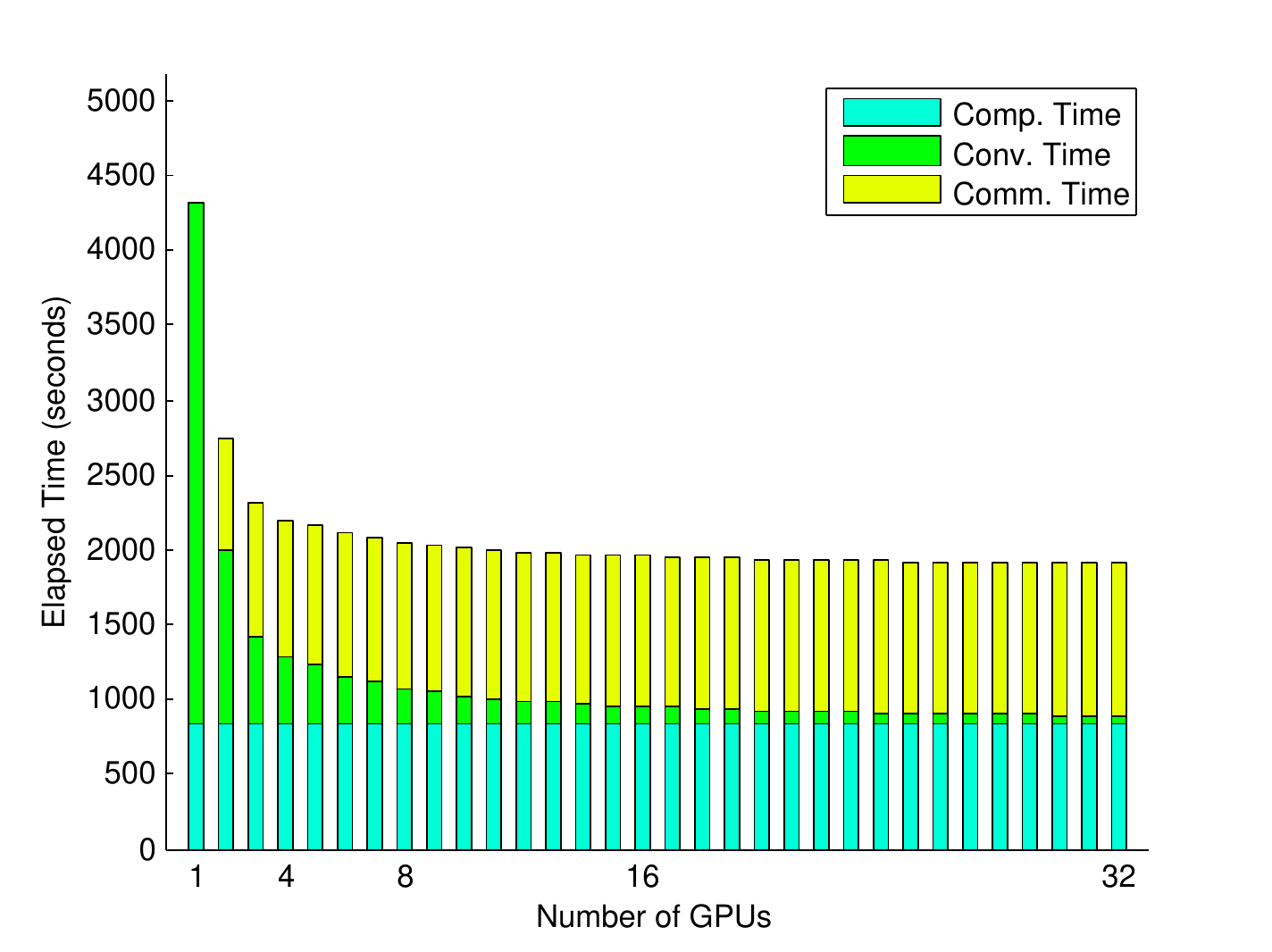}
		\caption{Elapsed time for the largest network, with a batch size of $1024$ images, using a GPU cluster ranging from $1$ to $32$ PCs.}
		\label{fig:Net_5-500-1500-PC_32-GPU}
	\end{center}
\end{figure}

The final case refers to the \ac{GPU} case, where only the largest network was simulated up to $32$ nodes (Figure \ref{fig:Net_5-500-1500-PC_32-GPU}). This is justified with the fact that the most efficient use of the \ac{GPU} occurs with the largest network, trained with the highest batch size of 1024 images. As in the \ac{CPU} case, the added nodes were considered to have computational capabilities similar to the devices used (shown in Table~3), being assigned performance values between worst and best \ac{GPU} case scenario used in the experiment. The results are detailed in Figure \ref{fig:Net_5-500-1500-PC_32-GPU}.

As in the cases considered for simulation using \acp{CPU}, the solution using \acp{GPU} is also scalable, with the speedup virtually stagnating for $8$ or more nodes. Since the convolution is done more quickly on a \ac{GPU} than on a \ac{CPU}, communication and computation times assume greater impact as bottlenecks. As stated previously, the communication time can be diminished with a faster data transmission, while the computation time can be improved with parallelization.

The results provided by TensorFlow source code~\cite{multigpu} serve as a mean of comparison to the scalability of this method. Said results show that the average attained speedups range from $2.6\times$ up to $3.5\times$, respectively for $2$ up to $4$ GPUs. Although these results are substantially better, there are two aspects that should be addressed. First, the GPUs used for the training with TensorFlow were all part of the same machine that requires specific hardware configuration and presents limitations (same hardware, same GPUs, max number of GPUs limited by motherboard), thus the data transmission rate is considerably higher than using devices in different machines, physically separated by the network, which incurs in significantly lower communication times.
	
Secondly, TensorFlow parallelizes the entire network, whereas our distribution technique focuses on the convolutional layers, thus turning the training of the remaining layers a bottleneck. Considering the case where  communication times are virtually nonexistent (which is achievable with faster communications), the speedup obtainable with our method is about $4.3\times$, which is slightly better.

In the end, our solution scales better to larger datasets with larger images, that require more convolutions.

\subsection{Discussion}

Despite achieving considerable speedups for the tested networks, some essential aspects of the experiment are to be considered.

The most important aspect to take into consideration is that speedups depend on the dataset and architecture used. This means that the best speedups achieved should be analyzed under the perspective of the CIFAR-10 dataset case. Thus, with an architecture more reliant on convolutions (more convolutional layers, or larger images), the amount of time spent during the convolutional phase is larger, leading to better speedups when using this type of distributed approach.

The next aspect to highlight is the difference between \ac{CPU} and \ac{GPU} speedups. This is explained by the fact that speedups are calculated with respect to the use of a single device of the same type. Since the convolutional layer is computed significantly faster with a \ac{GPU}, the computation phase will occupy a larger percentage in total processing time, thus decreasing the maximum speedup achievable.

Another aspect is Internet speed over distinct computing nodes, where data transmission rate is bound to vary. This impacts communication times, that influences final processing time and speedup. The approach used was conservative in the sense that the compute nodes have communicated exploiting low bandwidth rates, namely using Wi-Fi networks.

One important consideration pertains the used devices. Some of the laptops used are over two years old, which makes both the \acp{CPU} and \acp{GPU} considered low to mid range devices by today's standards. Using more recent devices from similar/same generation wouldn't differentiate much the results, since the used GPUs all belong to the same technology node, producing near maximum throughput performances in the range $790 \sim 1170$ GFLOPS. For the same reason, changing the base node of comparison among them wouldn't alter speedups significantly.

So, as a way to generalize even more the proposed distribution technique, both data transmission speed and performance values of the devices are varied and two cases considered. The first uses a cluster with both \acp{CPU} and \acp{GPU} for low to mid range devices, while the second case uses high-end devices on the cluster.

\begin{figure}[!h]
	\centering
	\begin{subfigure}[b]{0.47\textwidth}
		\includegraphics[width=\textwidth]{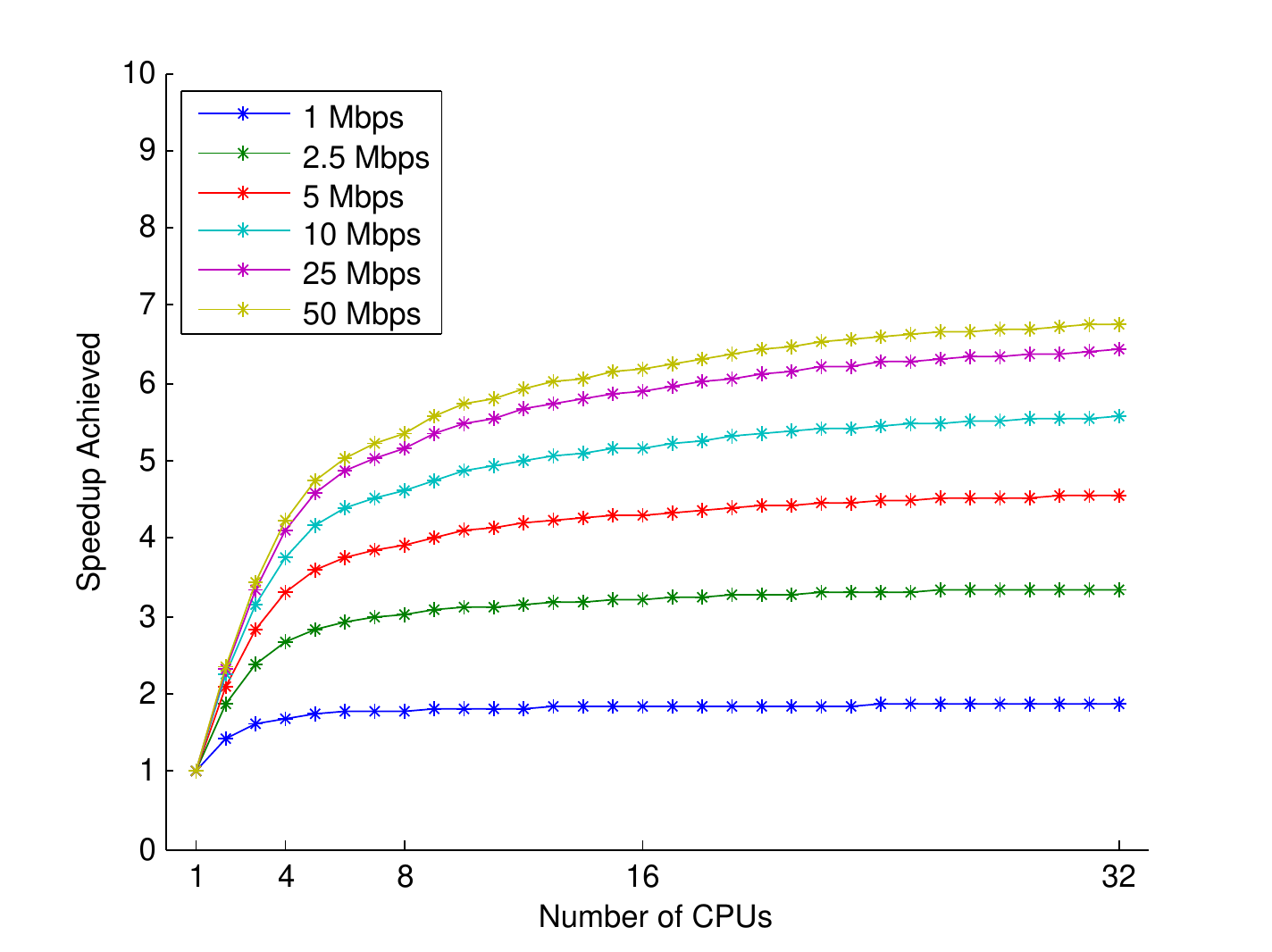}
		\caption{Low to mid range \ac{CPU} cluster}
		\label{fig:largeNetwork}
	\end{subfigure}
	\quad
	\begin{subfigure}[b]{0.47\textwidth}
		\includegraphics[width=\textwidth]{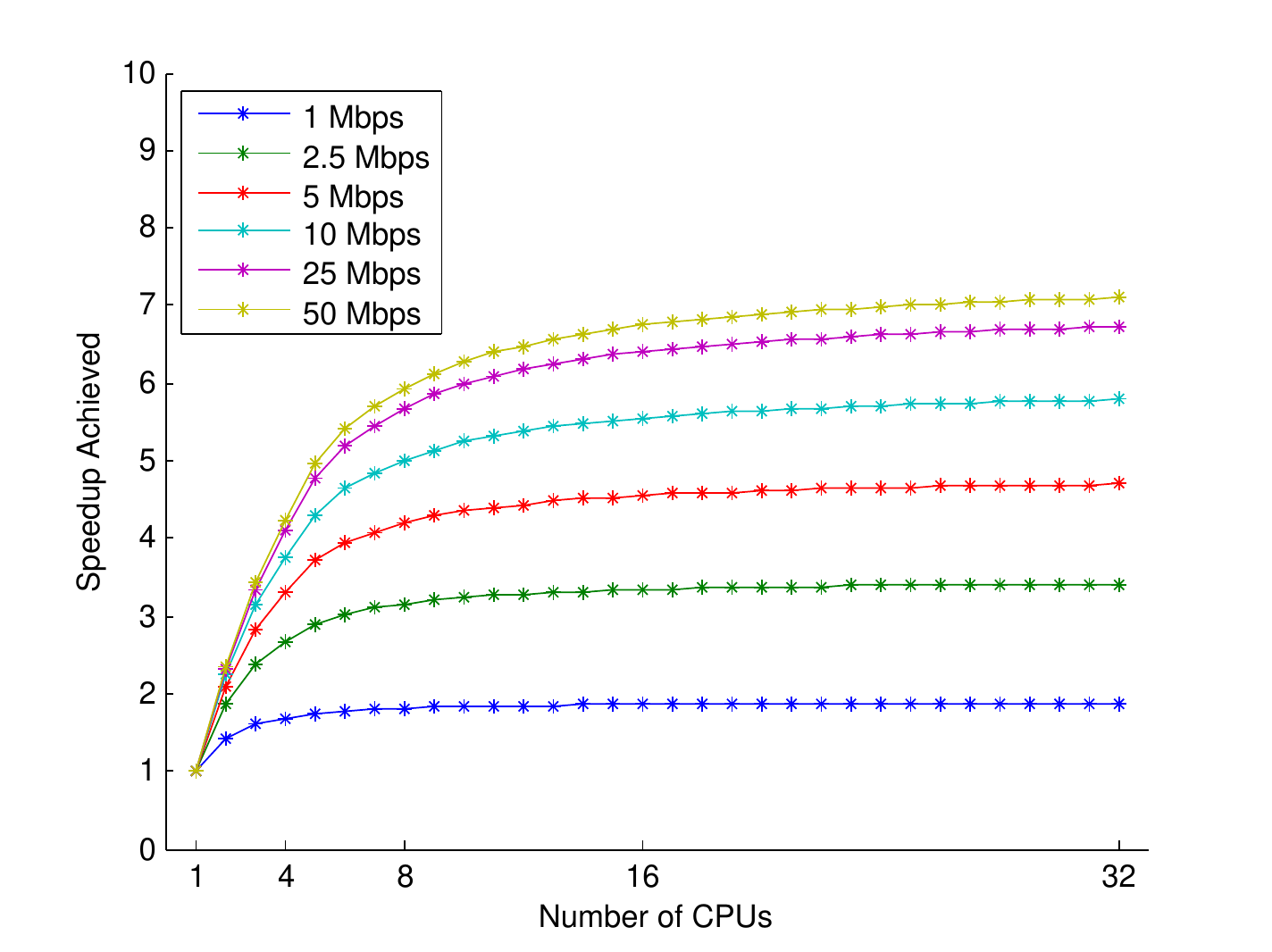}
		\caption{High-end \ac{CPU} cluster}
		\label{fig:goodpclargenetwork}
	\end{subfigure}
	\caption{Speedups achieved on the largest network, trained with $1024$ images for a cluster of $32$ nodes using a) low to mid range and b) high-end \acp{CPU}.}
	\label{fig:largenetwork}
\end{figure}
\begin{figure}[!h]
	\centering
	\begin{subfigure}[b]{0.47\textwidth}
		\includegraphics[width=\textwidth]{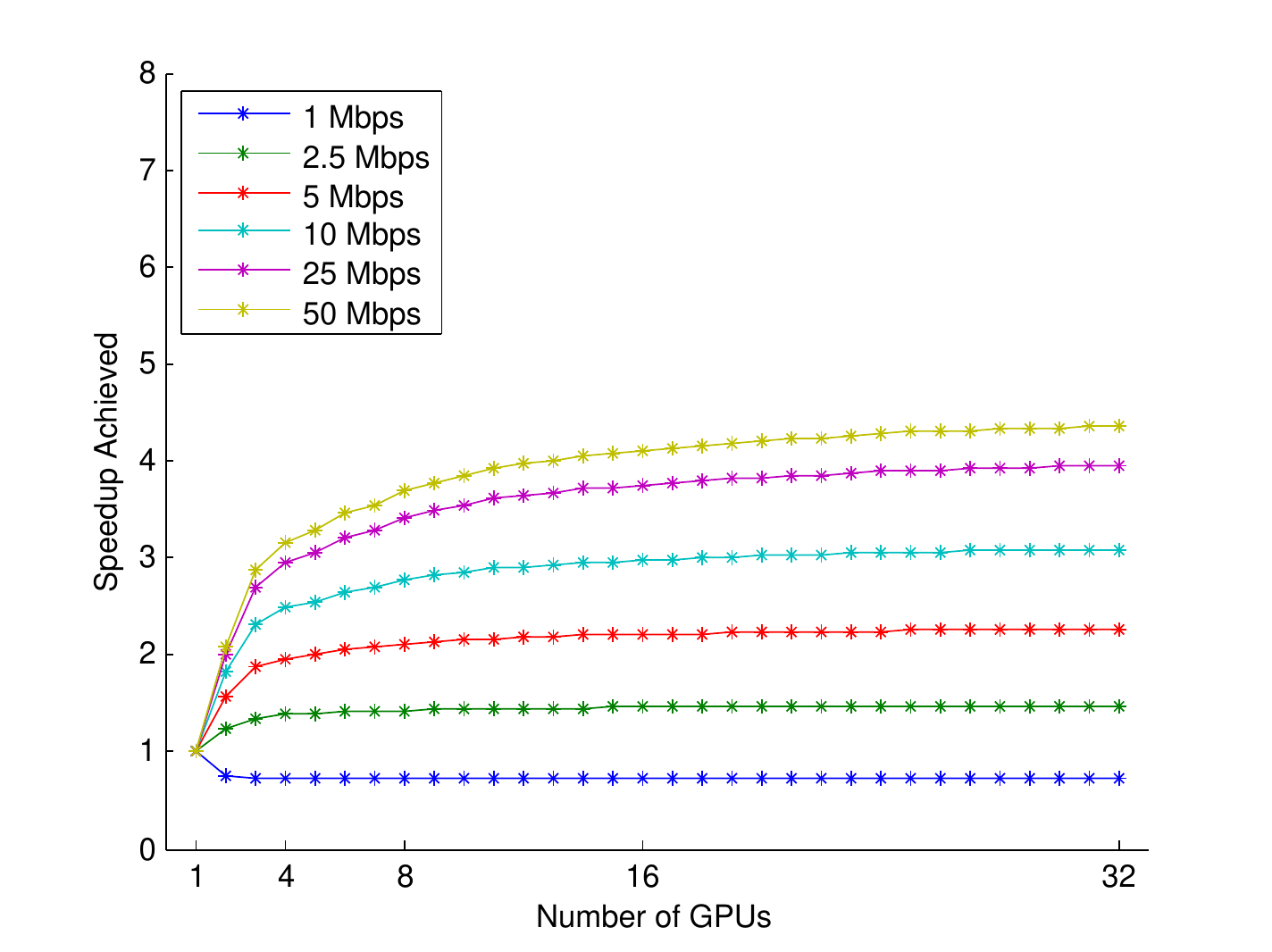}
		\caption{Low to mid range \ac{GPU} cluster}
		\label{fig:badGPU}
	\end{subfigure}
	\quad
	\begin{subfigure}[b]{0.47\textwidth}
		\includegraphics[width=\textwidth]{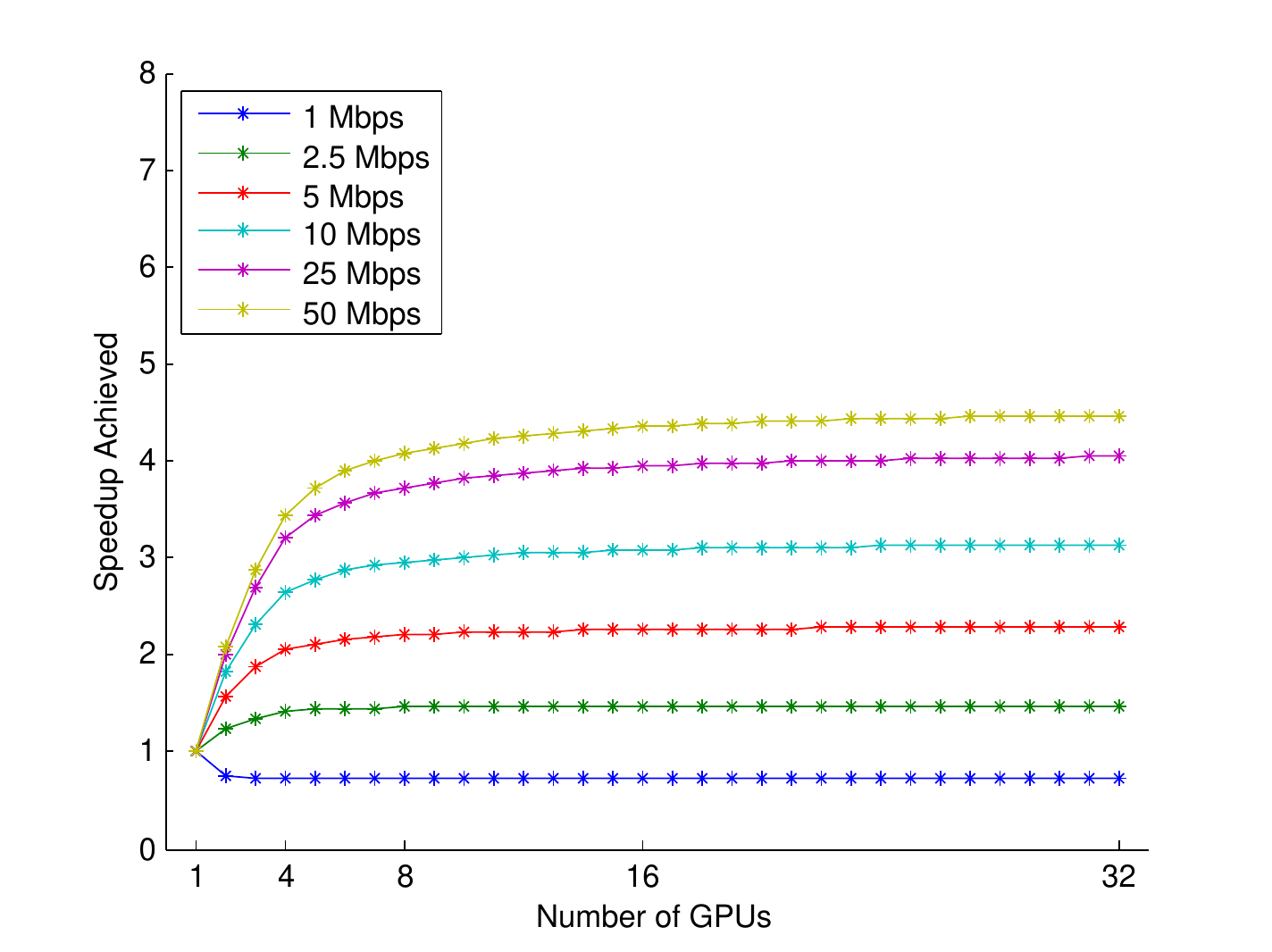}
		\caption{High-end \ac{GPU} cluster}
		\label{fig:goodGPU}
	\end{subfigure}
	\caption{Speedups achieved on the largest network, trained with $1024$ images for a cluster of $32$ nodes using a) low to mid range and b) high-end \acp{GPU}.}
	\label{fig:largenetworkGPU}
\end{figure}

The results for the \ac{CPU} and \ac{GPU} cases are presented in Figures \ref{fig:largenetwork} and \ref{fig:largenetworkGPU}, respectively.

Interestingly, maximum speedups achieved show that the difference between using low-end or high-end devices is negligible. This happens because the bottlenecks of this distribution technique end up being communication and computation time. This means that the only difference between using the two types of devices has to do with how many nodes are needed for the speedup to start stabilizing around a maximum, with fewer nodes required for the high-end devices.

However, Internet speed is extremely important, and this is justified by the fact that with faster data transmissions, the communication time stops acting as the main cause for a bottleneck and the network has the ability to achieve higher speedups. The contrary also stands true, with a slower data transmission diminishing the speedup, with the \ac{GPU} case showing that training may become even slower than using only $1$ \ac{GPU}.

Another aspect that deserves reflection is that the reported speedups preserve Amdhal's Law, indicating values below the theoretical maximum $10$ (which results from the acceleration of $60 \sim 90$\% of the global workload, i.e., convolution kernels). Also, the scalability of the proposed solution seems to slowdown for a number of nodes slightly above $10$, which acts inline with other works from the literature that fixed this value around $16$~\cite{keuper2016distributed}.

\subsubsection{Low-power and mobile GPUs}

\begin{figure}[!h]
	\centering
	\begin{subfigure}[b]{0.47\textwidth}
		\includegraphics[width=\textwidth]{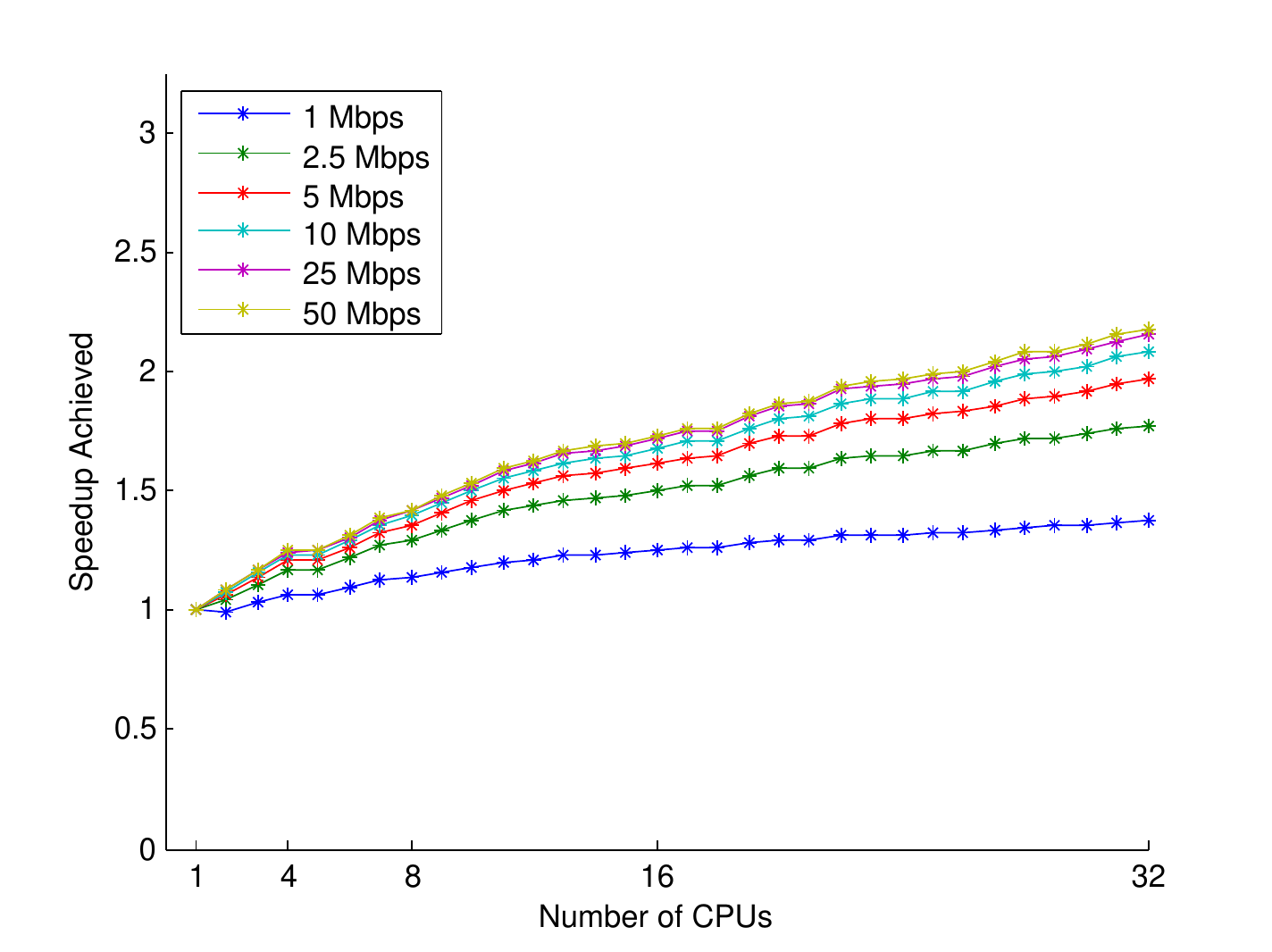}
		\caption{Mobile \ac{GPU} cluster using 32 nodes.}
		\label{fig:pc32largenetwork}
	\end{subfigure}
	\quad
	\begin{subfigure}[b]{0.47\textwidth}
		\includegraphics[width=\textwidth]{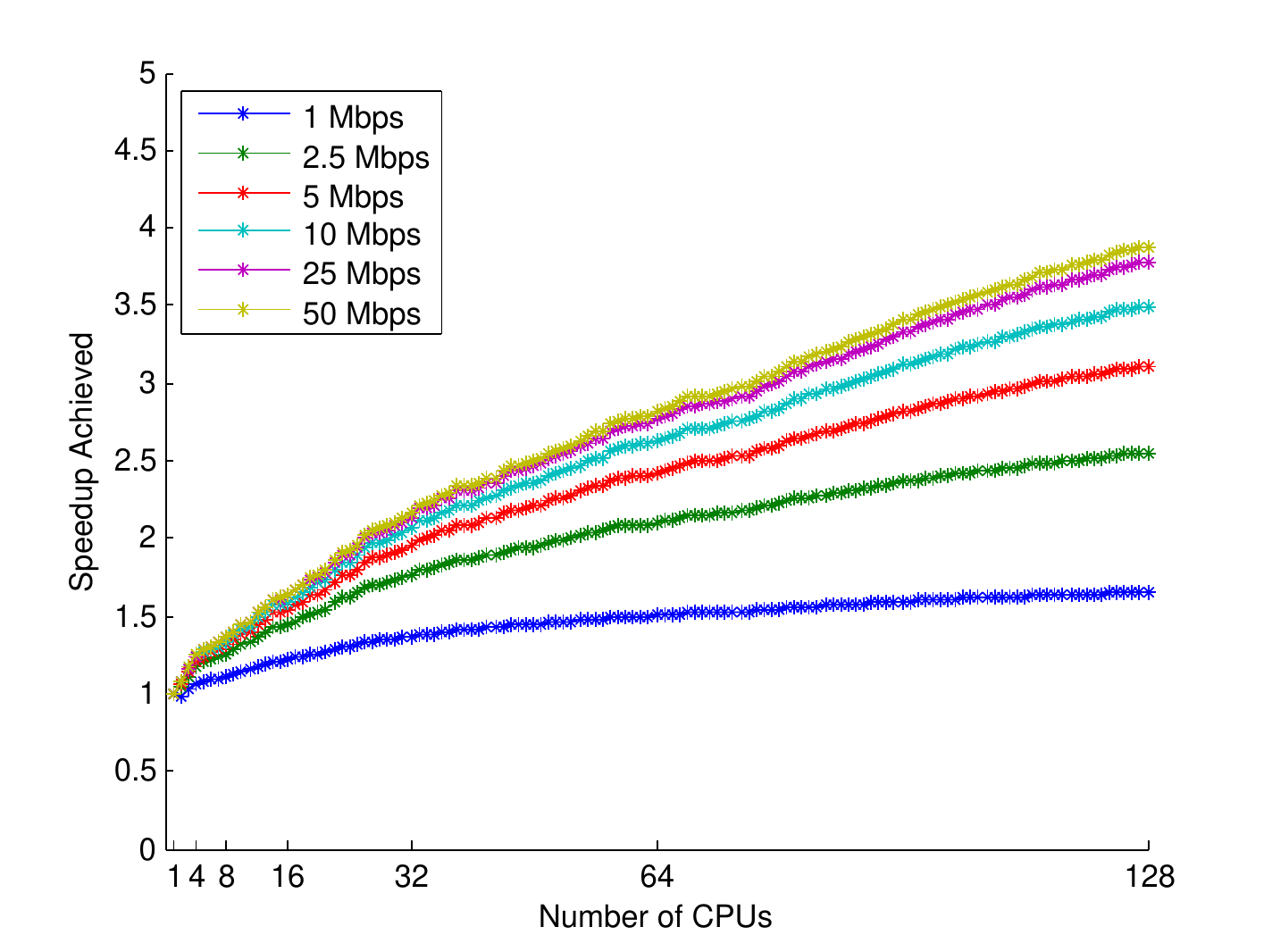}
		\caption{Mobile \ac{GPU} cluster using 128 nodes.}
		\label{fig:pc128largenetwork}
	\end{subfigure}
	\caption{Speedup achieved on the largest network, trained with 1024 images for a mobile \ac{GPU} cluster of a) 32 and b) 128 nodes.}
	\label{fig:shpc}
\end{figure}

Another aspect to consider consists of using devices that require less power but are slower than the ones used in these experiments. Specifically, mobile GPUs are about $10$ times slower~\cite{falcao2016evaluation} than the desktop \acp{GPU} used. For this simulation, the same variables as before are manipulated: Internet speed and number of nodes. One particular difference is that the master node was still a desktop \ac{GPU}. The results for this simulation are shown in Figure \ref{fig:shpc}.

An initial simulation only considered 32 nodes, but as subfigure \ref{fig:pc32largenetwork}) shows, 32 mobile \acp{GPU} are not sufficient to provide the same order of speedups that desktop \acp{GPU} do. Therefore new simulations were run using a maximum of 128 nodes. These results are documented in subfigure \ref{fig:pc128largenetwork}).

Although mobile \acp{GPU} have only a tenth of the processing power as their desktop counterpart and achieve considerably worse throughput performance, they should still be considered as a viable alternative, in particular because of power consumption requirements, with mobile GPUs achieving the same classification performance as the best GPUs in the market, but taking ten times longer~\cite{falcao2016evaluation} to complete execution. However, since their average power is nearly three orders of magnitude lower than the reference \ac{GPU}, it results in energy consumption around two orders of magnitude lower for the same amount of computation.

\section{Conclusions}

Unlike the methods of model parallelism, the technique proposed in this work only deals with the convolutional layer. Thus the only data that needs to be transmitted are the inputs and number of kernels, calculated during runtime, so that each node can compute its part of the balanced workload, which eliminates the convolution phase as a bottleneck, for implementations using both \acp{CPU} and \acp{GPU}.

That could be achieved with all the tested networks, with attained speedups for every architecture trained with the considered batch sizes. The best reached speedup using \acp{CPU} is 3.28x for 4 nodes, when training the largest network, with 1024 images in the batch, and is 2.45x for 3 \acp{GPU}, speedups that are completely dependent on the dataset used and architecture chosen (2 convolutional layers intertwined with 2 pooling layers). However, these speedups could be largely surpassed using faster data transmission, as the results from the simulations show. Even considering the case where only 2 devices are used (both \ac{CPU} and \ac{GPU}), the speedup always exists, and is close to 2x, on both tested cases.

Furthermore, this is the best technique to use with \acp{CPU} and/or \acp{GPU} with different processing resources, thanks to the capability of exploring hybrid \ac{CPU}-\ac{CPU} and \ac{GPU}-\ac{GPU} computation. Using data parallelism, the training time is always dependent on the slowest device, or in the case of asynchronism, the slowest device might train with old parameters. The alternative would be to split the data batch unevenly, but that would cause loss of information during the averaging of the parameters. With model parallelism, since it is necessary to define which neurons must communicate between nodes, it would be required to know all the processing information of each device \textit{a priori}.

The simulations further show that the attained speed\-ups depend very little of the processing capabilities of each individual device, for laptops ranging from low and mid range to high-end, for a number larger than 4, since the lower computational resources are largely compensated by the parallelization.

The decision of developing this method in Matlab was made not only due to the principle of generic adoption by science communities, or its intuitive nature regarding matrices and operations involving their calculations, but mainly to the possibility of working using devices with different operating systems, without the need to develop cross-platform software, which can become a complex and tedious task.

The developed solution proves to be a useful tool for the distributed training of \acp{CNN}. Although good performances were achieved, there is one other aspect that could, and should, be further explored, and that is implementation using \ac{OpenCL}, as opposed to \ac{CUDA}. Not only would that mean that other \acp{GPU} could be used, such as AMD's, but more importantly, it would allow for the distribution of the training to be done using mobile \acp{GPU}, as well as \acp{FPGA} and other low-power devices. Despite not having the same computational resources as desktop \acp{CPU} and \acp{GPU}, they can be far more energy efficient, and would allow to achieve smaller energy consumption levels without compromising the desired throughput and classification performance.

\section*{Acknowledgements}
This work was partially supported by Funda\c{c}\~{a}o para a Ci\^{e}ncia e a Tecnologia (FCT) and Instituto de Telecomunica\c{c}\~{o}es under grant UID/EEA/50008/2013.


\begin{thebibliography}{00}

\bibitem{lecun1997reading}
LeCun, Y., Bottou, L., Bengio, Y.: Reading checks with graph transformer
  networks.
\newblock In: International Conference on Acoustics, Speech, and Signal
  Processing, vol.~1, pp. 151--154. IEEE, Munich (1997)

\bibitem{kononenko2001machine}
Kononenko, I.: Machine learning for medical diagnosis: History, state of the
  art and perspective.
\newblock Artif. Intell. Med. \textbf{23}(1), 89--109

\bibitem{mnih2013atari}
Mnih, V., Kavukcuoglu, K., Silver, D., Graves, A., Antonoglou, I., Wierstra,
  D., Riedmiller, M.: Playing atari with deep reinforcement learning.
\newblock In: NIPS Deep Learning Workshop (2013)

\bibitem{kabir2015machine}
Kabir, M.H., Hoque, M.R., Seo, H., Yang, S.H.: Machine learning based adaptive
  context-aware system for smart home environment.
\newblock International Journal of Smart Home \textbf{9}, 55--62 (2015)

\bibitem{dixit2014use}
Dixit, A., Naik, A.: Use of prediction algorithms in smart homes.
\newblock International Journal of Machine Learning and Computing \textbf{4} (2014)

\bibitem{falcao2016evaluation}
Falcao, G., Alexandre, L.A., Marques, J., Frazao, X., Maria, J.: On the
  evaluation of energy-efficient deep learning with stacked autoencoders on
  mobile gpus (2017).
\newblock {25th Euromicro International Conference on Parallel,
  Distributed, and Network-Based Processing} (2017)

\bibitem{krizhevsky2012imagenet}
Krizhevsky, A., Sutskever, I., Hinton, G.E.: Imagenet classification with deep
  convolutional neural networks.
\newblock In: F.~Pereira, C.J.C. Burges, L.~Bottou, K.Q. Weinberger (eds.)
  Advances in Neural Information Processing Systems 25, pp. 1097--1105. Curran
  Associates, Inc. (2012)
  
\bibitem{corrado2012large}
Dean, J., Corrado, G.S., Monga, R., Chen, K., Devin, M., Le, Q.V., Mao, M.Z.,
  Ranzato, M., Senior, A., Tucker, P., Yang, K., Ng, A.Y.: Large scale
  distributed deep networks.
\newblock In: NIPS (2012)
  
\bibitem{OpenCL}
Stone, J.E., Gohara, D., Shi, G.: Opencl: A parallel programming standard for
  heterogeneous computing systems.
\newblock IEEE Des. Test \textbf{12}(3), 66--73 (2010)

\bibitem{CUDA}
{NVIDIA Corporation}: NVIDIA CUDA Compute Unified Device Architecture
  Programming Guide.
\newblock NVIDIA Corporation (2007)
  
\bibitem{ward2011efficient}
Ward, J., Andreev, S., Heredia, F., Lazar, B., Manevska, Z.: Efficient mapping of the training of convolutional neural networks to a cuda-based cluster (2011)
  
\bibitem{krizhevsky2014weird}
Krizhevsky, A.: One weird trick for parallelizing convolutional neural
  networks.
\newblock CoRR \textbf{abs/1404.5997} (2014)
  
\bibitem{gene2013computer}
Amdahl, G. M.: Computer Architecture and Amdahl's Law.
\newblock IEEE Computer. \textbf{46}(12), 38--46 (2013)

\bibitem{keuper2016distributed}
Keuper, J., Pfreundt, F-J.: Distributed training of deep neural networks: theoretical and practical limits of parallel scalability.
\newblock In: 2nd Workshop on Machine Learning in HPC Environments (2016)

\bibitem{erhan2013scalable}
Erhan, D., Szegedy, C., Toshev, A., Anguelov, D.: Scalable object detection
  using deep neural networks.
\newblock CoRR \textbf{abs/1312.2249} (2013)

\bibitem{szegedy2013deep}
Szegedy, C., Toshev, A., Erhan, D.: Deep neural networks for object detection.
\newblock In: C.J.C. Burges, L.~Bottou, M.~Welling, Z.~Ghahramani, K.Q.
  Weinberger (eds.) Advances in Neural Information Processing Systems 26, pp.
  2553--2561. Curran Associates, Inc. (2013)
  
\bibitem{toth2014combining}
Toth, L.: Combining time- and frequency-domain convolution in convolutional
  neural network-based phone recognition.
\newblock In: 2014 IEEE International Conference on Acoustics, Speech and
  Signal Processing (ICASSP), pp. 190--194 (2014)
  
\bibitem{abdelhamid2014convolutional}
Abdel-Hamid, O., Mohamed, A.R., Jiang, H., Deng, L., Penn, G., Yu, D.:
  Convolutional neural networks for speech recognition.
\newblock IEEE/ACM Trans. Audio, Speech and Lang. Proc. \textbf{22}(10), 1533--1545 (2014)

\bibitem{lai2015giraffe}
Lai, M.: Giraffe: Using deep reinforcement learning to play chess.
\newblock CoRR \textbf{abs/1509.01549} (2015)

\bibitem{silver2016mastering}
Silver, D., Huang, A., Maddison, C.J., Guez, A., Sifre, L., van~den Driessche,
  G., Schrittwieser, J., Antonoglou, I., Panneershelvam, V., Lanctot, M.,
  Dieleman, S., Grewe, D., Nham, J., Kalchbrenner, N., Sutskever, I.,
  Lillicrap, T., Leach, M., Kavukcuoglu, K., Graepel, T., Hassabis, D.:
  Mastering the game of {Go} with deep neural networks and tree search.
\newblock Nature \textbf{529}(7587), 484--489 (2016)

\bibitem{Fukushima1980}
Fukushima, K.: Neocognitron: A self-organizing neural network model for a
  mechanism of pattern recognition unaffected by shift in position.
\newblock Biological Cybernetics \textbf{36}(4), 193--202 (1980)

\bibitem{lecun1995convolutional}
LeCun, Y., Bengio, Y.: Convolutional networks for images, speech, and
  time-series.
\newblock In: M.A. Arbib (ed.) The Handbook of Brain Theory and Neural
  Networks. MIT Press (1995)
  
\bibitem{lecun1989backpropagation}
LeCun, Y., Boser, B., Denker, J.S., Henderson, D., Howard, R.E., Hubbard, W.,
  Jackel, L.D.: Backpropagation applied to handwritten zip code recognition.
\newblock Neural Comput. \textbf{1}(4), 541--551 (1989)

\bibitem{ranzato2013multi}
Yadan, O., Adams, K., Taigman, Y., Ranzato,M.: Multi-gpu training of convnets.
\newblock CoRR \textbf{https://arxiv.org/abs/1312.5853} (2013)

\bibitem{tensorflow2015}
Abadi, M., Agarwal, A., Barham, P., et~al.: {TensorFlow}: Large-scale machine
  learning on heterogeneous systems (2015).
\newblock Software available from tensorflow.org

\bibitem{krizhevsky2009learning}
Krizhevsky, A.: Learning multiple layers of features from tiny images.
\newblock Tech. rep., University of Toronto (2009)

\bibitem{multigpu}
Abadi, M., Agarwal, A., Barham, P., et~al.: cifar10\_multi\_gpu\_train.py
  (2015).
\newblock \url{https://github.com/tensorflow/tensorflow/blob/0.6.0/tensorflow/models/image/cifar10/cifar10_multi_gpu_train.py}

\bibitem{vishnu2016distributed}
Vishnu, A., Siegel, C., Daily, J.: Distributed tensorflow with {MPI}.
\newblock CoRR \textbf{abs/1603.02339} (2016).

\bibitem{torralba2008tiny}
Torralba, A., Fergus, R., Freeman, W.T.: 80 million tiny images: A large data
  set for nonparametric object and scene recognition.
\newblock IEEE Trans. Pattern Anal. Mach. Intell. \textbf{30}(11), (1958--1970)

\bibitem{ILSVRC15}
Russakovsky, O., Deng, J., Su, H., Krause, J., Satheesh, S., Ma, S., Huang, Z.,
  Karpathy, A., Khosla, A., Bernstein, M., Berg, A.C., Fei-Fei, L.: {ImageNet
  Large Scale Visual Recognition Challenge}.
\newblock International Journal of Computer Vision (IJCV) \textbf{115}(3),
  211--252 (2015)

\end{thebibliography}


\end{document}